\documentclass[final,11p,a4paper]{elsarticle}

\journal{Future Generation Computer Systems}
\usepackage{color}
\usepackage[usenames,dvipsnames]{xcolor}

\usepackage{multicol}
\usepackage{booktabs}
\usepackage{siunitx}
\usepackage{textcomp}
\usepackage{pst-func}
\usepackage{amssymb}
\usepackage{graphicx}

\usepackage{paralist}
\usepackage{environ}
\usepackage{pgfplots}
\usepackage{tikz}
\usepackage{forest}

\usetikzlibrary{shadows}
\usetikzlibrary{pgfplots.dateplot}
\usetikzlibrary{matrix,chains,positioning,decorations.pathreplacing}
\usetikzlibrary{patterns,calc,fit,arrows}
\usetikzlibrary{shapes.geometric}

\usepackage{mathtools}
\usepackage{array,multirow}
\usepackage{algorithm}
\usepackage[hyphens]{url}

\usepackage{pgfplotstable}
\usepackage{filecontents}

\usepackage{mdframed}

\usepackage{arydshln}
\usepackage{etoolbox}

\usepackage{xspace} % Smart spaces after commands
 
\usepackage{dirtytalk} % for \say command
 
\usepackage{makecell}% for multirow in table header
\usepackage{stackengine}
\setstackEOL{\cr} %EOL is abbreviation for "end of line."

\usepackage[inline,shortlabels]{enumitem} 
\newlist{inlinelist}{enumerate*}{1}
\setlist*[inlinelist,1]{%
  label=(\roman*),
}

\usepackage{nicefrac}

\usepackage[hidelinks]{hyperref}
\usepackage{cleveref}
\usepackage{bigints}

\usepackage{showkeys}%to show labels

\usepackage{pdflscape}
\usepackage{pgfplotstable}
\usepackage{filecontents}

\usetikzlibrary{pgfplots.dateplot}
\usetikzlibrary{matrix,chains,positioning,decorations.pathreplacing}
\usetikzlibrary{patterns,calc,fit,arrows}
\usetikzlibrary{shapes.geometric}

\usepackage[final,nomargin,inline,index]{fixme} % Simplified management of FIXME's
\fxusetheme{color}

\FXRegisterAuthor{bart}{anbart}{\color{magenta} {\underline{bart}}}
\FXRegisterAuthor{tizi}{antizi}{\color{red} {\underline{tizi}}}

\hypersetup{
  breaklinks   = true,
  colorlinks   = true, %Colours links instead of ugly boxes
  urlcolor     = blue, %Colour for external hyperlinks
  linkcolor    = blue, %Colour of internal links
  citecolor    = red   %Colour of citations
}
\hypersetup{final} % Needed to generate hyperlinks even in draft mode (ERROR IN MAC)

\usepackage{listings}
\definecolor{listingBG}{HTML}{FFFFCB}%
\definecolor{listingFrame}{HTML}{BBBB98}%
\definecolor{listingLineno}{rgb}{0.5,0.5,1.0}%
\definecolor{LightGrey}{rgb}{0.975,0.975,0.975}
\lstdefinelanguage{solidity}{
	commentstyle=\color{Gray},
	morecomment=[l]{//},
	morecomment=[s]{/*}{*/},
	classoffset=0,
        escapechar=\$,
	morekeywords={struct,mapping,function,this,public,private,static,final,class,extends,switch,case,break,finally,try,catch,return,if,else,new},
	keywordstyle=\color{Blue}\bfseries,
	classoffset=1,
	morekeywords={unit,int,string,bool,address,uint},
	keywordstyle=\color{TealBlue},
	classoffset=2,
	morekeywords={ether,wei,finney,contract,send,throw,msg,sender,value},
	keywordstyle=\color{Plum}\bfseries,
}

\lstdefinelanguage{java}{
	escapechar=\$,
        commentstyle=\color{Gray},
	morecomment=[l]{//},
	morecomment=[s]{/*}{*/},
	morestring=[b]",
        classoffset=0,
	morekeywords={public,private,static,final,class,extends,switch,case,break,finally,try,catch,void,int,boolean,throws,throw,return,if,else,new},
	keywordstyle=\color{keyword}\bfseries
}

\usepackage{refcount}

\newcommand{\myfootnotetext}[1]{\footnotetext{#1\label{fn:text}%
        \edef\fnmark{\getpagerefnumber{fn:mark}}%
        \edef\fntext{\getpagerefnumber{fn:text}}%
        \ifx\fnmark\fntext\else\ClassWarning{}{footnote mark and text on different pages!}\fi}}

\newcommand{\codefont}{\fontsize{10}{10}\selectfont}
\newcommand{\code}[1]{{\tt {#1}}}

\def\etc{\emph{etc}.\@\xspace}

\newcommand{\eg}{e.g.\@\xspace}
\newcommand{\ie}{i.e.\@\xspace}

\newcommand{\wei}{\textit{wei}\xspace}
\newcommand{\ether}{\textit{ETH}\xspace}
\newcommand{\USD}{\mbox{\textit{USD}}\xspace}
\newcommand{\USDfmt}[1]{\SI[round-precision=0,round-mode=places]{#1}}

\newmdenv [linewidth=0pt]{mdNoFramed}
\newcommand{\lineno}[1]{{\tt\codefont {\textcolor{magenta}{#1}}}}

\makeatletter
\pgfplotsset{
    /pgfplots/flexible xticklabels from table/.code n args={3}{%
        \pgfplotstableread[#3]{#1}\coordinate@table
        \pgfplotstablegetcolumn{#2}\of{\coordinate@table}\to\pgfplots@xticklabels
        \let\pgfplots@xticklabel=\pgfplots@user@ticklabel@list@x
    }
}
\makeatother

\def\gainList{{Government,EthereumPyramid,ProtectTheCastle,TreasureChest,ZeroPonzi,Ethstick,Doubler2,DynamicPyramid},
              {Etheramid1,LittleCactus,GreedPit,Doubler,ShinySquirrels1,Thesimplegame,Newponzi,Doubler3},
              {Quick1,Myscheme,Bunny,DoubleTx,Multi33v,DianaEthereum-x18,Rubixi}}

\newcounter{gainListCounter}

\newcommand{\urldataset}{\href{http://goo.gl/CvdxBp}{\code{goo.gl/CvdxBp}}\xspace}
\newcommand{\urlgithub}{\href{https://github.com/blockchain-unica/ethereum-ponzi}{\code{github.com/blockchain-unica/ethereum-ponzi}}\xspace}

\newcommand{\ethlifetimeyears}{3\xspace}

\newcommand{\gldavg}{0.79\xspace}
\newcommand{\gldthreshold}{0.35\xspace}

\newcommand{\avgPubLifetime}{393\xspace} % updated 2019-01-25
\newcounter{valTotTx}
\newcounter{valPubPonzi}
\newcounter{valHidPonzi}
\newcounter{valHidVerPonzi}
\newcounter{valQuasiPonzi}
\newcounter{valTotPonzi}
\newcommand{\pubPonzi}{\arabic{valPubPonzi}\xspace}

\newcommand{\hidVerPonzi}{\arabic{valHidVerPonzi}\xspace}
\newcommand{\quasiPonzi}{\arabic{valQuasiPonzi}\xspace}
\newcommand{\totPonzi}{\arabic{valTotPonzi}\xspace}

\newcommand{\totInTx}{18925\xspace}
\newcommand{\totOutTx}{9100\xspace}
\newcommand{\totInETH}{\SI[round-precision=0,round-mode=places]{43881}[]{}}
\newcommand{\totOutETH}{\SI[round-precision=0,round-mode=places]{43332}[]{}}
\newcommand{\totInUSD}{\SI[round-precision=0,round-mode=places]{630662}[]{}}
\newcommand{\totOutUSD}{\SI[round-precision=0,round-mode=places]{702878}[]{}}
\newcommand{\totInUsr}{2378\xspace}
\newcommand{\totOutUsr}{1232\xspace}

\newtoggle{forest}
\toggletrue{forest}

\newtoggle{draft}
\togglefalse{draft}

\makeatletter
\renewcommand\paragraph{\@startsection{paragraph}{4}{\z@}%
  {2.25ex \@plus 1ex \@minus .2ex}%
  {-0.75em}%
  {\normalfont\normalsize\bfseries}}
\makeatother

\begin{document}

\begin{frontmatter}

\title{Dissecting Ponzi schemes on Ethereum: \\ identification, analysis, and impact\tnoteref{ack}}
\tnotetext[ack]{Work partially supported by Aut.\ Region of Sardinia under projects ``Sardcoin'', ``Smart Collaborative Engineering'', and P.I.A.\ 2013 ``NOMAD''.}

\author{Massimo Bartoletti\corref{bart}}
\author{Salvatore Carta\corref{a2}}
\author{Tiziana Cimoli\corref{a3}}
\author{Roberto Saia\corref{a4}}

\cortext[bart]{\emph{Corresponding author}. Dipartimento di Matematica e Informatica, Universit\`a degli Studi di Cagliari, via Ospedale 72, 09124 Cagliari (Italy), e-mail: \texttt{bart@unica.it}}

\address{Dipartimento di Matematica e Informatica, Universit\`a degli Studi di Cagliari, Italy}

\begin{abstract}
  Ponzi schemes are financial frauds which lure users
  under the promise of high profits.
  Actually, users are repaid only with the investments of new users joining the scheme:
  consequently, a Ponzi scheme implodes soon after users stop joining it.
  Originated in the offline world 150 years ago, 
  Ponzi schemes have since then migrated to the digital world,
  approaching first the Web,
  and more recently hanging over cryptocurrencies like Bitcoin.
  Smart contract platforms like Ethereum have provided 
  a new opportunity for scammers,
  who have now the possibility of creating ``trustworthy''
  frauds that still make users lose money, 
  but at least are guaranteed to execute ``correctly''.
  We present a comprehensive survey of Ponzi schemes on Ethereum,
  analysing their behaviour and their impact from various viewpoints.
\end{abstract}

\begin{keyword}
smart contracts \sep cryptocurrencies \sep Ponzi schemes \sep electronic frauds
\end{keyword}

\end{frontmatter}

\section{Introduction}
\label{sec:intro}

The advent of Bitcoin~\cite{bitcoin,Bonneau15ieeesp} has given birth to a new way to exchange currency, 
allowing secure and (almost) anonymous transfers of money 
without the intermediation of trusted authorities.
This has been possible by suitably combining several techniques, 
among which digital signature schemes, moderately hard ``proof-of-work'' puzzles,
and the idea of \emph{blockchain}, an immutable public ledger which records all the money transactions,
and is maintained by a peer-to-peer network through a distributed consensus protocol.

Soon after Bitcoin has become widespread, 
it has started arousing the interest of criminals,
eager to find new ways to transfer currency 
without being tracked by investigators and surveillance authorities~\cite{Moore13ijcip}.

Recently, \emph{Ponzi schemes}~\cite{Artzrouni09} ---
a classic fraud originated in the offline world at least 150 years ago ---
have approached the digital world, 
first on the Web~\cite{Moore12fc}, 
and more recently also on Bitcoin~\cite{Vasek15fc}.
Ponzi schemes are often disguised as ``high-yield'' investment programs.
Users enter the scheme by investing some money.
The actual conditions which allow to gain money depend on the specific rules of the scheme,
but all Ponzi schemes have in common that, 
to redeem their investment, one has to make new users enter the scheme.
A more authoritative definition of Ponzi schemes
comes from the U.S.\ Securities and Exchange Commission (SEC):%
\footnote{Source: \href{https://www.sec.gov/spotlight/enf-actions-ponzi.shtml}{\code{www.sec.gov/spotlight/enf-actions-ponzi}}.}

\label{def:ponzi-sec}

\begin{center}
\begin{minipage}{0.95\textwidth}
\say{{\it%
A Ponzi scheme is an investment fraud that involves the payment of purported returns to existing investors from funds contributed by new investors. Ponzi scheme organizers often solicit new investors by promising to invest funds in opportunities claimed to generate high returns with little or no risk. With little or no legitimate earnings, Ponzi schemes require a constant flow of money from new investors to continue. Ponzi schemes inevitably collapse, most often when it becomes difficult to recruit new investors or when a large number of investors ask for their funds to be returned.%
}}
\end{minipage}
\end{center}

Often, the investment mechanism of Ponzi schemes 
creates a pyramid-shape topology of users, 
having at the top level the initiator of the scheme,
and at level $\ell+1$ the users who compensate the investment of those at level $\ell$.
The scheme will eventually collapse
because at some point it will no longer be possible to find new investors,
as their number grows exponentially in the number of levels of this pyramid.
Therefore, users at the top levels of the pyramid will gain money,
while those at the bottom levels will just lose their investment.

Despite many investors are perfectly conscious of the fraudulent nature of these schemes,
and of the fact that they are illegal in many countries,
Ponzi schemes continue to attract remarkable amounts of money.
A recent study~\cite{Vasek15fc} estimates that
Ponzi schemes operated through Bitcoin have gathered
more than $7$ millions \USD in the period from September 2013 to September 2014%
\footnote{%
  This estimate considers both traditional Ponzi schemes which also accept payments in bitcoins,
  and schemes that only handle bitcoins.
}.

\paragraph{``Smart'' Ponzi schemes.}

The spread of \emph{smart contracts},
\ie,
computer programs whose correct execution is automatically 
enforced without relying on a trusted authority~\cite{Szabo97firstmonday},
creates new opportunities for fraudsters.
Indeed, implementing Ponzi schemes as smart contracts
would have several attractive features:
\begin{enumerate}

\item The initiator of a Ponzi scheme could stay anonymous, 
  since creating the contract and withdrawing money from it
  do not require to reveal her identity;
  
\item Since smart contracts are ``unmodifiable'' and ``unstoppable'', 
  no central authority (in particular, no court of law)
  would be able to terminate the execution of the scheme, 
  or revert its effects in order to refund the victims.
  This is particularly true for smart contracts running 
  on \emph{permissionless} blockchains, 
  which are controlled by a peer-to-peer network of nodes.

\item Investors may gain a false sense of trustworthiness from the
  fact that the code of smart contracts is public and immutable, and
  their execution is automatically enforced.  
  This may lead investors
  to believe that the owner cannot take advantage of their money, that
  the scheme would run forever, %
  and that they have a fair probability of gaining the declared interests.

\end{enumerate}

All these features are made possible by a combination of factors,
among which the growth of platforms for smart contracts~\cite{Lamela16iacr},
which advertise anonymity and contract persistence as main selling points,
and the fact that these technologies are very recent, 
and still live in a gray area of legal systems~\cite{Murphy15crs,Juels16ccs}.

Understanding the behaviour of ``smart'' Ponzi schemes 
would be crucial to devise suitable intervention policies.
To this purpose, one has to analyse various aspects of the fraud, answering several questions:
how many victims are involved? 
How much money is invested?
What are the temporal evolution and the lifetime of a fraud?
What kind of users fall in these frauds?
Can we recognize fingerprints of Ponzi schemes during their execution,
or possibly even before they are started?
Investigating on these issues would help to disrupt this kind of frauds.

  \paragraph{Contributions.}

This paper is the first comprehensive survey on Ponzi schemes 
in Ethereum~\cite{ethereum},
the most prominent platform for smart contracts so far.
We construct a dataset of Ponzi schemes,  
and we analyse them from various perspectives.
More specifically, our contributions can be summarised as follows:
\begin{itemize}

\item a set of criteria for determining when a smart contract
  implements a Ponzi scheme. 
  Our criteria take into account only the logic implemented by the contract
  to gather and distribute money, while neglecting external factors,
  like \eg, the way the scheme is advertised or gamified.

\item a public dataset of Ponzi schemes deployed on Ethereum
  (\urldataset),
  Coherently with our classification criteria,
  the dataset is constructed by examining the source code of contracts.
  We start from the contracts whose source code is available on blockchain explorers,
  finding among them \pubPonzi Ponzi schemes. 
  We expand this collection to \totPonzi schemes,
  by searching the blockchain for contracts whose bytecode is highly similar to a contract already 
  classified as Ponzi.
  False negatives are excluded by manually inspecting their decompiled code.

\item an open-source tool (\urlgithub) 
  which extracts from the Ethereum blockchain all the
  transactions of the Ponzi schemes in our dataset,
  records all the incoming and outgoing movements of money,
  and computes the analyses presented in this paper.

\item an analysis of the source code of the contracts in our collection 
  (\Cref{sec:source}).
  We discover that most contracts share a few common patterns,
  and that many of them are obtained by minor variations of 
  already existing ones. 
  We devise a rough taxonomy of Ponzi schemes, 
  which classifies them according to the pattern used to redistribute payouts.
  We show that the schemes in each category fail to achieve a fair distribution of money.
  Further, we spot several security vulnerabilities in the analysed contracts, 
  which could be exploited by adversaries to steal money.
  
\item a measure of the economic impact of Ponzi schemes, 
  quantifying the overall value exchanged through them 
  (\Cref{sec:impact}).

\item a measure of the gains and losses of the users of Ponzi schemes
  (\Cref{sec:gains-losses}).
  We focus on the top $10$ schemes
  (those with the highest number of transactions), 
  and on further $13$ schemes that we select among the ones 
  with most interesting features
  (number of users, transactions, or ether exchanged).  
  In most cases we observe the typical pattern of Ponzi schemes: 
  a few users gain a lot, while the majority of users simply
  lose their money.

\item an analysis of the temporal behaviour of Ponzi schemes 
  under various viewpoints 
  (\Cref{sec:timing}).
  First, we investigate the lifetime of Ponzi schemes, 
  an important indicator to predict when a scheme is going to collapse.
  Then, we analyse the correlation
  between inflow and outflow of contracts over time. 
  Finally, we measure the monthly volume of transactions.
  
\item a measure of the inequality of payments to and from the schemes
  (\Cref{sec:inequality}).
  This indicator may reveal how scammers select their victims:
  a fair distribution of payments means that the scheme is fed 
  by a large number of victims who pay small amounts of money;
  instead, an unequal distribution often means that the scheme profits 
  from a small number of ``big fishes'' who invest a lot of money.
  
\item a set of guidelines that users could follow to 
  protect themselves against Ponzi schemes
  (\Cref{sec:conclusions}).

\end{itemize}

\section{Ethereum in a nutshell}
\label{sec:ethereum}

Ethereum~\cite{ethereum} is a decentralized virtual machine, 
which can execute programs --- called \emph{contracts} ---
written in a Turing-complete bytecode language,
called EVM~\cite{ethereumyellowpaper}.
Every contract has a
permanent storage where to keep data, and a set of functions
which can be invoked either by users or by other contracts.  
Users and contracts can own a crypto-currency 
(called \emph{ether}, or $\ether$ in short),
and send/receive ether to/from users or other contracts.

Users can send \emph{transactions} to the Etherum network in order to:
\begin{inlinelist}
\item create new contracts;
\item invoke a function of a contract;
\item transfer ether to contracts or to other users.
\end{inlinelist}
All the transactions sent by users,
called \emph{external} transactions,
are recorded on a public, append-only
data structure --- the \emph{blockchain}.
Upon receiving an external transaction, 
a contract can fire some
\emph{internal} transactions,
which are not explictly recorded on the blockchain,
but still have effects on the balance of users and of other contracts.

Since transactions can move money, 
it is crucial to guarantee that their execution is performed correctly.
To this purpose, Ethereum does not rely
on a trusted central authority: rather, each transaction is processed
by a decentralized network of nodes.
There is a \emph{consensus} protocol to address mismatches 
(due \eg, to failures or to attacks), 
which is currently based on a ``proof-of-work'' puzzle. 
The security of the consensus protocol relies on the fact that
following the protocol is more convenient than trying to attack it.
Indeed, nodes receive  economic incentives for correctly performing
all the computations required by the protocol.  
The execution of contracts is guaranteed to be correct, as long
as the adversary does not control a very large portion of the computational
power of the network~\cite{Gervais16ccs}.

\paragraph{Contracts.}

Abstractly, contracts can be seen as objects in object-oriented languages,
which are composed of fields and functions.
A user can invoke a function by sending a suitable transaction 
to the Ethereum nodes.
The transaction must include the execution fee (for the miners), 
and may include a transfer of ether from the caller to the contract.

We illustrate contracts through a small example
(\code{AWallet}, in~\Cref{ex:wallet}),
which implements a personal wallet associated to an owner.
Rather than programming it directly as EVM bytecode, 
we use \emph{Solidity}, a Javascript-like programming language 
which compiles into EVM bytecode~\cite{solidity}.
The contract can receive ether from other users,
and its owner can send (part of) that ether to other users 
via the function \code{pay}.
The hashtable \code{outflow} records all the \emph{addresses}%
\footnote{Addresses are sequences of 160 bits which uniquely identify contracts and users.}
to which it sends money, and associates to each of them the total
transferred amount.  
The hashtable \code{inflow} records all the addresses
from which it has received money.
All the ether received is held by the contract.  
Its amount is automatically recorded in \code{balance}: 
this is a special variable, which cannot be altered by the programmer.  
When a contract receives ether, it also executes a special function 
with no name, called \emph{fallback} function.

\begin{figure}[t]
  \begin{center}
    \scalebox{1}{%
\begin{lstlisting}[language=solidity,numbers=left,numbersep=12pt]
contract AWallet{
    address owner;
    mapping (address => uint) public outflow;
    mapping (address => uint) public inflow;
    
    function AWallet(){ owner = msg.sender; }
    
    function pay(uint amount, address recipient) returns (bool){
        if (msg.sender != owner ||  msg.value != 0) throw;
        if (amount > this.balance) return false; 
	outflow[recipient] += amount;
        if (!recipient.send(amount)) throw;
        return true;
    }

    function(){ inflow[msg.sender] += msg.value; }
}
\end{lstlisting}
    } 
  \end{center}
  \vspace{-5pt}
  \caption{A simple wallet contract.}  
  \label{ex:wallet}
\end{figure}

The function \code{AWallet} at line~\lineno{6} is a constructor,
run only once when the contract is created.
The function \code{pay} sends \code{amount} \wei
($1\, \wei = 10^{-18} \ether$) from the contract to
\code{recipient}. 
At line~\lineno{9} the contract throws an exception if
the caller (\code{msg.sender}) is not the owner, or if some ether
(\code{msg.value}) is attached to the invocation and transferred to
the contract.
Since exceptions revert side effects, 
this ether is returned to the caller (who however loses the fee).
At line~\lineno{10}, the call terminates if the required amount
of ether is unavailable; in this case, there is no need to revert the
state with an exception.
At line~\lineno{11}, the contract updates the \code{outflow} registry,
before transferring the ether to the recipient. 
The function \code{send} used at line~\lineno{12} to this purpose
presents some quirks,
\eg it may fail if the recipient is a contract.
The fallback function at line~\lineno{16} is triggered upon receiving ether,
when no other function is invoked. 
In this case, the fallback function just updates the \code{inflow} registry. 
In both cases, when receiving ether and when sending, the total amount
of ether of the contract, stored in variable \code{this.balance}, 
is automatically updated.

\section{Collection of Ponzi schemes}
\label{sec:methodology}

In this~\namecref{sec:methodology}
we establish a set of criteria for classifying contracts as Ponzi schemes.
We then describe our methodology for constructing a collection of Ponzi schemes,
and for extracting the related transactions.

\subsection{What is a ``smart'' Ponzi scheme?}
\label{sec:methodology:traits}

We start by clarifying what is considered a Ponzi scheme in this paper.
The first key choice that we make is to restrict to
the schemes which are implemented as smart contracts ---
or \emph{``smart'' Ponzis}~\cite{Alphaville17}.
This choice rules out scams which use Ethereum 
only as a mean of payment (or just for advertisement).
These scams include some well-known ``high-yield'' investment programs,
many of which
are reported on the blacklist maintained by \emph{BadBitcoin}%
\footnote{\href{https://badbitcoin.org/thebadlist/index.php}{\code{https://badbitcoin.org/thebadlist/index.php}}}.
We chose to exclude this kind of scams from our analysis, since
it is seldom possible to retrieve any information about 
the Ethereum addresses they use (if any).

In~\Cref{fig:ponzi-classification} we propose four requirements
to determine if a contract is a Ponzi scheme,
based exclusively on the logic implemented within the contract.
When a contract satisfies \emph{all} four requirements, 
we classify it as a Ponzi scheme.

\begin{figure}[t]
\fbox{
\begin{minipage}{\textwidth}
\medskip
\begin{description}

  \item[R1] the contract  distributes money among investors, according to some logic.

  \item[R2] the contract receives money \emph{only} from investors.

  \item[R3] each investor makes a profit \emph{if} enough investors invest enough money in the contract afterwards.

  \item[R4] the later an investor joins the contract, the greater the risk of losing his investment.

\end{description}
\medskip
\end{minipage}
}
\caption{Criteria for classifying a contract as a Ponzi scheme.}
\label{fig:ponzi-classification}
\end{figure}

\begin{itemize}

\item \textbf{R1} asks that the contract 
distributes money to \emph{investors}, \ie users who join the contract by sending some money to it.
This requirement does not put any constraints on the logic used to distribute the money, 
so \textbf{R1} alone is not enough to classify a contract as Ponzi:
for instance, gambling games, lotteries, insurances and bonds, satisfy \textbf{R1}.
However, \textbf{R1} rules out the contracts which provide users with some kind of assets,
but do not implement the logic to distribute them:
rather, these assets are exchanged through external marketplaces,
like cryptocurrency exchanges.
This is the case, \eg, of most implementations of 
ERC-20 tokens~\cite{MaxwellERC20},
among which Initial Coin Offerings~\cite{Holoweiko19ico}.

\item \textbf{R2} asks that the money gathered by the contract
comes from investors, \emph{only}.
This rules out the cases where the money distributed to investors comes from
external sources, like \eg a bank which pays the interests of 
a ``smart'' bond,
or a bookmaker who pays off bets using his own funds.

\item \textbf{R3} asks that each investor makes a profit,
\emph{provided that} 
new investors continue to send money to the contract.
Together with the first two requirements, this implies that
users make profits \emph{only} through the investments of other users.
Note that gambling games, betting, and lotteries violate \textbf{R3}:
there, even if there is a constant flow of investments,
an unlucky user is not guaranteed to make any profit
(\eg, he can always lose the lottery).

\item \textbf{R4} asks that the risk
of losing one's investment grows with the time one joins the scheme.
This is a landmark feature Ponzi schemes also in the real world:
at a certain point it becomes difficult to find new investors, 
so no one makes profits anymore, and the scheme collapses.

\end{itemize}

Compared to the SEC definition of Ponzi scheme quoted in~\Cref{sec:intro},
the requirements \textbf{R1} and \textbf{R2} together capture that fact that a Ponzi scheme
\emph{``involves the payment of purported returns to existing investors from funds contributed by new investors''};
\textbf{R3} corresponds to the fact that they \emph{``require a constant flow of money from new investors to continue''};
\textbf{R4} implies that they \emph{``inevitably collapse, most often when it becomes difficult to recruit new investors''}.
Note that our requirements do not capture the fact that
\emph{``Ponzi scheme organizers often solicit new investors by promising to invest funds in opportunities claimed to generate high returns with little or no risk''}.
This is because, by design, our requirements are based exclusively on the logic implemented within the contract,
while advertising is done outside the contract code.

\paragraph{Ponzi \emph{vs.} pseudo-Ponzi schemes}

Note that requirement \textbf{R4} rules out some contracts which are 
sometimes blamed to be Ponzi schemes, even if the 
Ponzi mechanism to distribute investments 
is not \emph{hard-coded} in the contract.
This is the case \eg of contracts which implement 
crypto-collectible markets ---
the most notable instance being \emph{CryptoKitties},
a game where players can breed and trade virtual cats,
implemented as ERC-721 tokens.
Indeed, \textbf{R4} is violated, because a lucky user,
regardless of the moment when he joins the contract,
may breed a rare cat, and make a profit by its sale.

For similar reasons, \textbf{R4} rules out \emph{Fomo3D},
a sort of game which is sometimes pointed out as a Ponzi scheme.
\emph{Fomo3D} works as a lottery game where, at each round, 
players can purchase some ``keys'', and the last buyer in the round
wins a jackpot.
Whenever a key is purchased, the deadline to the end of the round
is extended, and the earning from key selling is split in two parts: 
a part is added to the jackpot, while
the other is shared among the participants in the round.
The lottery mechanism decouples the time when a user joins the scheme
from its risk of losing her investment, violating \textbf{R4}.

Requirement \textbf{R4} rules out also \emph{PoWH3D},
another alleged Ponzi scheme.
\emph{PoWH3D} implements a token and its exchange:
the contract mechanism ensures that
the value of the token grows when people buy, 
and decreases when they sell;
further every token trade has a $10\%$ fee.
Investors can earn in three ways:
by selling a token for more than it was paid;
by inviting a new investor to buy tokens 
(in this case, they get the fees of the invitee);
and by receiving the fees paid by a (not invited) investor
(these fees are distributed among all the token holders).
Requirement \textbf{R4} is violated because
investing late, \eg in a period of stagnation, does not necessarily 
imply a greater risks of losing one's investment,
since the mechanism ensures that the value of tokens is low.

We remark that, even if a contract does not explicitly implement
a Ponzi mechanism (so, violating some of our requirements),
it may potentially induce a behaviour 
which closely resembles that of a Ponzi scheme.
For instance, \emph{CryptoKitties} and its followers
gave rise to a market of crypto-collectibles which
is often compared to the ``tulip mania'', 
a large speculative bubble in the 1600s.
The extreme popularity of \emph{CryptoKitties}
almost caused the congestion of the Ethereum network in 2017;
some virtual cats were sold for more than $170 K \USD$,
and the market has processed more than \$12 million 
in sales of virtual cats~\cite{Young17cointelegraph}.

\subsection{Collection of Ponzi schemes}
\label{sec:methodology:solidity}

To construct a dataset of Ponzi schemes, 
we start by retrieving the Solidity code of contracts
published on the Ethereum blockchain.
Since the blockchain only stores the EVM bytecode,
to this purpose we rely on the blockchain explorer \emph{Etherscan},
which allows developers to upload the Solidity code of their contracts, 
and verifies that their compilation matches the EVM code on the 
blockchain%
\footnote{When we first created our collection in 2017, 
it was still possible to list all contracts with verified source code
through the URL
\url{https://etherscan.io/contractsVerified}.
Currently, only the last 500 contracts with verified source are listed.
To overcome this limitation, one can use a 
blockchain parser, like \eg BlockAPI~\cite{Bartoletti17serial},
to scan all the transactions on the blockchain,
and fetch their Solidity code from \emph{Etherscan}.
For a contract address \code{xyz}, the URL \url{https://etherscan.io/address/xyz\#code} contains the contract Solidity code,
if verified by \emph{Etherscan}.}.

By manually inspecting the Solidity code of these contracts, 
we detect \pubPonzi contracts which satisfy \emph{all} 
the requirements in~\Cref{fig:ponzi-classification},
and therefore can be classified as Ponzi schemes.
Since all the contracts in this sample are relatively small 
($<120$ LOC, including comments), 
manual inspection was accurate enough to check the requirements.
As a further check, for all these contracts we study 
the pattern they use to redistribute money,
which is the basis for our taxonomy in~\Cref{sec:source}.
To stay on the safe side (\ie, to avoid false positives),
we have not included in this collection those contracts which 
are too complex to establish with certainty
whether they satisfy the requirements or not.

We perform a second search phase to enlarge our collection.
More specifically, we search the Ethereum blockchain for contracts 
whose bytecode is \emph{similar} to that of some Ponzi scheme
identified in our initial collection.
This is done through the following steps:
\begin{enumerate}

\item We use a Monte Carlo algorithm to estimate the 
\emph{normalized Levenshtein distance}~\cite{ngld} (NLD) 
between two \emph{arbitrary} EVM contracts on the Ethereum blockchain.
The NLD is a standard measure of similarity between two strings.
The \emph{non-normalized} Levenshtein distance between two strings 
measures the number of character which one has to change to transform the first string in the second one
(\eg, the distance between ``Ponzi'' and ``Banzai'' is 3).
The \emph{normalized} version is a metric, 
and its value is a real number ranging between 0 (perfect equality) 
and 1 (perfect inequality).
After these calculations, we estimate as \gldavg
the NLD between two arbitrary EVM contracts downloaded from the blockchain.

\item We compute the NLD between the contracts in our initial sample, 
and all the contracts on the Ethereum blockchain.
We consider as a \emph{potential} Ponzi scheme any contract with a NLD 
less than \gldthreshold from some contract in our sample.
The two values \gldthreshold and \gldavg are sufficiently far apart
to ensure a low incidence of \emph{false positives},
\ie contracts whose NLD from the initial sample is below \gldthreshold,
but they are not Ponzi schemes.
This search resulted in \quasiPonzi \emph{potential} new Ponzi schemes,
not included in our original collection of \pubPonzi contracts.

\item We apply the Online Solidity Decompiler%
\footnote{\url{https://ethervm.io/decompile}}
to the EVM bytecode of the
\quasiPonzi contracts found in the second phase,
and we manually compare the obtained Solidity code with that
of the corresponding Ponzi scheme found in the first phase.
In \hidVerPonzi cases we find a substantial match between the contract codes,
so we add these contracts to our collection.

\end{enumerate}

In conclusion, we end up with a dataset of \totPonzi Ponzi schemes,
which we make available at \urldataset
(an excerpt is in~\Cref{fig:ponzi-stats} in~\Cref{sec:impact}).
We stress that our collection does \emph{not} include all the Ponzi schemes
which have been published on Ethereum over the years.
For instance, the contract \code{PonziUnlimited}%
\footnote{\url{https://etherscan.io/address/0x582b2489710A4189AD558B6958641789587fCc27}}
is blatantly a Ponzi scheme,
but it is not immediate to detect if its logic 
satisfies the requirements in~\Cref{fig:ponzi-classification},
so we do not include it in our collection%
\footnote{A relevant question, without an easy answer, would be that
of estimating the total number of smart Ponzi schemes on Ethereum.
The analysis in~\cite{Chen19access} conjectures that there could be 507 
smart Ponzi schemes created on Ethereum before September 2017.}.

\subsection{Extraction of transactions}
\label{sec:methodology:transactions}

For each Ponzi scheme in our dataset,
we gather all its transactions (both external and internal) 
from the Ethereum blockchain.
More specifically, for each transaction we record the following data: 
\begin{inlinelist}
\item the number of the enclosing block;
\item the date when it was published on the blockchain;
\item the address of the sender;
\item the address of the receiver;
\item the amount of ether transferred by the transaction;
\item a boolean value which records whether the transaction execution resulted in an error;
\item a boolean value which indicates whether the transaction is external or internal.
\end{inlinelist}
The scripts that we have developed to this purpose
exploit the Etherscan Ethereum Developer APIs%
\footnote{\url{https://etherscan.io/apis}},
and they are available at~\urlgithub.

\section{Anatomy of Ponzi schemes}
\label{sec:source}

In this~\namecref{sec:source} we analyse the source code of Ponzi schemes,
to understand their behaviour, and find analogies between different schemes. 
We then discuss some security issues found in the analysed contracts.

\subsection{Taxonomy of Ponzi schemes}

Based on the analysis of the contracts source code carried out in~\Cref{sec:methodology:solidity},
we devise a rough taxonomy of Ponzi schemes,
which classifies them according to the pattern used to redistribute money.
Our taxonomy consists of four categories, whose archetypal representatives
are displayed in~\Cref{ex:tree-shaped,ex:chain-shaped,ex:waterfall,ex:handover}%
\footnote{The code snippets presented there assume version v0.2.2 of the
Solidity compiler, which is the version used by most of the contracts in our collection.
Although more recent versions of Solidity change the way to declare functions and to manage arrays, 
these changes do not really affect the spirit of our examples.}. 
We discuss below the categories of our taxonomy.

\paragraph{Tree-shaped schemes} 
use a tree data structure to induce an ordering among users.  
Whenever a user joins the scheme, she must indicate another user as inviter, 
who becomes her {\it parent node}.  
If no inviter is indicated, the parent will be the root node, \ie the owner of the scheme.  
In most schemes, the amount of money to be invested is chosen by the user, 
and there is a lower bound to that amount.  
The money of the new user is split among her ancestors
with the logic that the nearest an ancestor is, the greater her share.
Since there is no limit to the number of children of a node, 
the more children (and descendants) a node has, the more money it will make.

We show in~\Cref{ex:tree-shaped} an archetypal scheme of this kind.
To join the scheme, a user must send some money, 
and must indicate an \code{inviter} that will be her parent node. 
If the amount is too low (line~\lineno{15}), 
or if the user is already present  (line~\lineno{16}), 
or if the inviter does not exist (line~\lineno{17}), the user is rejected;
otherwise she is inserted in the tree (line~\lineno{19}). 
Once the user has joined, her
investment is shared among her ancestors (lines~\lineno{25-29}),
halving the amount  at each level.
 
\begin{figure}[t]
  \begin{minipage}[t]{0.45\textwidth}
\begin{lstlisting}[language=solidity,lastline=15,numbers=left,numbersep=10pt]
contract TreePonzi {

  struct User {
    address inviter;
    address itself;
  }
  mapping (address=>User) tree;
  address top;

  function TreePonzi() {
    tree[msg.sender] =
      User({itself: msg.sender,
           inviter:msg.sender});
    top = msg.sender;
  }
\end{lstlisting}
  \end{minipage}
  \hspace{10pt}
  \begin{minipage}[t]{0.65\textwidth}
\begin{lstlisting}[language=solidity,firstline=17,firstnumber=16,numbers=left,numbersep=10pt]
  function enter(address inviter) public {
    if ((msg.value < 1 ether) ||
      (tree[msg.sender].inviter != 0x0) ||
      (tree[inviter].inviter == 0x0)) throw;
         
    tree[msg.sender] = User({itself: msg.sender,
	                     inviter: inviter});
    address current = inviter;
    uint amount = msg.value;   
    while (next != top) {
      amount = amount/2;
      current.send(amount);
      current = tree[current].inviter;
    }
    current.send(amount);
  }}
\end{lstlisting}
  \end{minipage}    
  \vspace{-15pt}
  \caption{A tree-shaped scheme.}  
  \label{ex:tree-shaped}
\end{figure}

In this scheme, a user cannot foresee how much she will gain: this
depends on how many users she is able to invite, and on how much they
will invest.  
The only one who is guaranteed to have profit is the owner, 
\ie the root node of the tree.  
Examples for this kind of scheme are \code{Etheramid} and \code{DynamicPyramid}.

\paragraph{Chain-shaped  schemes} are a special case of tree-shaped schemes, 
where each node of the tree has exactly one child
(so, the ordering induced among users is linear). 
The schemes in this category usually multiply the investment by 
a predefined constant factor, which is equal for all users.
The scheme starts paying back users, one at a time, in order of arrival, and in full:
all new investements are gathered until the due amount is obtained.
At that moment, the contract sends the payout back in a single shot,
and moves on to the next user in the chain.
The amount to be invested can be fixed, or free, or have a lower bound. 
Usually, the contract owner retains a fee from each investment. 

We show in~\Cref{ex:chain-shaped} an archetypal chain-shaped scheme,
which doubles the investment of each user. 
To join the scheme, a user sends \code{msg.amount} \ether to the contract, 
hence triggering the fallback function (line~\lineno{14}).
The contract requires a minimum fee of $1$\ether: 
if \code{msg.amount} is below this minimum, the user is rejected (line~\lineno{15});
otherwise, her address is added to the array (line~\lineno{17}),
and the array length is incremented%
\footnote{In Solidity, \href{http://solidity.readthedocs.io/en/develop/types.html\#members}{dynamic arrays}
  can be resized by changing the \code{length} member.}.
The contract owner retains 10\% of the investment (line~\lineno{22}).
With the remaining funds, the contract tries to pay back the previous users.
If the \code{balance} is enough to pay the user at index \code{paying}, 
the contract pays the user her investment multiplied by 2 (line~\lineno{25}). 
After that, the contract tries to pay the next user, 
and so on until the balance is enough.

In this scheme, a user can foresee exactly how much she will gain, 
provided that the scheme keeps running;
the amount is proportional to what she has invested.  
Examples of this kind are \code{Doubler}, \code{DianaEthereum}, and \code{ZeroPonzi}.

\begin{figure}
  \begin{minipage}[t]{0.4\textwidth}
\begin{lstlisting}[language=solidity,lastline=14,numbers=left,numbersep=10pt]
contract ChainPonzi {

  struct User {
    address addr;
    uint amount;
  }
  User[] public users;
  uint public paying = 0;
  address public owner;
  uint public totalUsers=0;

  function ChainPonzi() {
    owner = msg.sender;
  }
\end{lstlisting}
  \end{minipage}
  \hspace{5pt}
  \begin{minipage}[t]{0.75\textwidth}
\begin{lstlisting}[language=solidity,firstline=15,firstnumber=15,numbers=left,numbersep=10pt]

  function() {
    if (msg.value < 1 ether) throw;

    users[users.length] = User({addr: msg.sender,
                                amount: msg.value});
    totalUsers += 1;
    owner.send(msg.value/10);

    while (this.balance > users[paying].amount * 2) {
      users[paying].addr.send(users[paying].amount * 2);
      paying += 1;
    }
  }}
\end{lstlisting}
  \end{minipage}
  \vspace{-15pt}
  \caption{A chain-shaped scheme.}  
  \label{ex:chain-shaped}
\end{figure}

\paragraph{Waterfall schemes} are similar to chain shaped-schemes for 
the user ordering, yet  different for the logic of money distribution.  
Each new investment is poured along the chain of investors, so that each can take their share.
Since the logic is first-come first-served, 
and the distribution starts always from the beginning of the chain, 
the users later in the chain are likely to never get any money.

We show in~\Cref{ex:waterfall} an archetypal scheme of this kind,
with an entry toll of $1$\ether (line~\lineno{19}), 
$10\%$ fees for the owner (line~\lineno{24}), and
a payout of $6\%$ of user investments at each turn.
The payout logic starts at line~\lineno{27}.
If the contract \code{balance}
is enough to pay the first user in the array (at position \code{pos = 0}), 
then the contract sends to that user $6\%$ of her original investment 
(lines~\lineno{29-30}). 
After that, the contract tries to pay the next user in the array, and
so on until the balance is exhausted.  
On the subsequent investment, the array is iterated again, still starting from the first user.

To ensure that all users receive payouts (coherently with requirement \textbf{R3}), 
the investments of new users must grow proportionally to the number of users.
Examples for this kind of scheme are \code{TreasureChest} and \code{PiggyBank}.

\begin{figure}
  \begin{minipage}[t]{0.4\textwidth}
\begin{lstlisting}[language=solidity,lastline=17,numbers=left,numbersep=10pt]
contract WaterfallPonzi {

  struct User {
      address addr;
      uint amount;
  }

  User[] public users;

  uint pos = 0;
  uint public totalUsers=0;
  address public owner;
  uint public fees = 0;
  
  function WaterfallPonzi() {
    owner = msg.sender;
  }
\end{lstlisting}
  \end{minipage}
  \hspace{10pt} 
  \begin{minipage}[t]{0.65\textwidth}
\begin{lstlisting}[language=solidity,firstline=19,firstnumber=18,numbers=left,numbersep=10pt]
  function() {
    if (msg.value < 1 ether) throw;
    
    users[totalUsers] = User({addr: msg.sender,
	                     amount:msg.value});
    totalUsers += 1;
    fees = mgs.value / 10;
    owner.send(fees);
   
    pos=0;	
    while (this.balance >= users[pos].amount*6/100 && pos<totalUsers){                              
      users[pos].etherAddress.send
	(users[pos].amount * 6/100);                      
      pos += 1; 
}}}
\end{lstlisting}
  \end{minipage} 
  \vspace{-15pt}
  \caption{A waterfall scheme.}  
  \label{ex:waterfall}
\end{figure}

\paragraph{Handover schemes} are an instance of chain-shaped scheme, 
where the entry toll is determined by the contract, 
and it increased each time a new investor joins the scheme.
The toll of a new investor is given {\it in full} to the previous one:
since the entry toll is increasing, the previous investor makes an instant profit.
At each moment, there is only one investor who is receiving money, 
and as soon as she is paid, she hands that privilege over the next user.  

An archetypal example is shown in~\Cref{ex:handover}. 
To join the scheme a user must send at least \code{price} \ether to the contract, 
hence triggering the fallback function of line~\lineno{11}. 
The contract forwards that sum to the former \code{user}, 
minus a fee  which is kept within the contract (line~\lineno{13}).  
Then, the address of the new user is recorded (line~\lineno{14}), 
and the \code{price} is doubled  (line~\lineno{15}). 
The contract owner can withdraw her share by calling \code{sweepCommission}.

\begin{figure}
  \begin{minipage}[t]{0.4\textwidth}
\begin{lstlisting}[language=solidity,lastline=10,numbers=left,numbersep=10pt]
contract HandoverPonzi {
  address owner;
  address public user; 
  uint public
    price = 100 finney;

  function HandoverPonzi() {
    owner = msg.sender;
    user = msg.sender;
  }
\end{lstlisting}
  \end{minipage}
  \hspace{10pt}
  \begin{minipage}[t]{0.65\textwidth}
\begin{lstlisting}[language=solidity,firstline=12,firstnumber=11,numbers=left,numbersep=10pt]
  function() {
    if (msg.value < price) throw;
    user.send(msg.value * 9 / 10);
    user = msg.address;
    price = price * 2;
  }

  function sweepCommission(uint amount)  {
    if (msg.sender == owner) owner.send(amount);
  }}
\end{lstlisting}
  \end{minipage}
  \vspace{-15pt}
  \caption{An handover scheme.}  
  \label{ex:handover} 
  % \noindent\;
\end{figure}

In handover schemes, at the time of the investment users know exactly how much they will gain. 
However, since the toll increases as the scheme goes on, 
later users are more likely to lose their money (coherently with requirement \textbf{R4}).
A paradigmatic representative of handover schemes is \code{KingOfTheEtherThrone}.

\subsection{ Analysis of money redistribution}

Ponzi schemes have the peculiarity that
each investor can make a profit, provided that  \emph{enough} investors invest
\emph{enough} money in the contract, after him
(requirement \textbf{R3}).
Focussing on the kinds of schemes identified in the previous section,
we now study how many users an investor must wait for, and how much they must invest to make her (say) double her money.

\paragraph{Chain-shaped  schemes}
Consider a chain-shaped scheme which {\it doubles} the received money, 
accepts entry tolls of exactly 1\ether, and has no owner fees except the first
1\ether sent to the contract.
Let us assume that the first user $U_1$ sends $1$\ether.  
Her money is given to the owner, and so it is removed from the contract,
whose balance is $0$.  
For $U_1$ to see back her 1\ether plus the other one
promised (since the contract doubles the investment),
 he must wait for two others users $U_2$ and $U_3$ to join
the scheme, by sending 1\ether each.

In~\Cref{fig:ponziTree}, each node represents one user, 
and its children are the users needed to redeem her share. 
So, $U_2$ must wait for $U_1$ to redeem her share, 
and then he must wait for $U_3$ and $U_4$ to send money 
(hence he has to wait a total of 3 users). 
User $U_3$, who is the last one on her level, must wait that the 
subsequent level is full, which gives a total of 4 users to wait.
In general, a user $U_k$ at level $i$ must wait that all those
users on the previous level have redeemed their share, 
and then he must wait for all the ones on her level 
that have arrived before him.
If $U_k$ is the first node at level $i$, he must
wait for all the other users at level $i$ to join, 
plus the two ones needed to redeem her share. 
This needs $2^{i} -1 + 2$ users. 
Since the amount of nodes up to level $i-1$ is $2^{i} -1$ and since $U_k$ is the first at level $i$, 
we have that $ k = 2^i$ and hence, in the best case, $U_k$ must wait $k+1$ users.
Instead, if $U_k$ is the last user at level $i$, he must
wait for all the other users at level $i+1$. 
This needs $2^{i+1}$ users. 
Since, $k = 2^{i+1} -1$, in this case $U_k$ must wait for $k+1$ users.
For instance, a user joining the scheme at position $3$ must wait $4$ new users, 
and to invest $4$\ether, in order to double her investment; 
instead, a user at position $50$ must wait for $51$ new users.

Although this simple example considers a scheme 
with no fees and a fixed investment for each user,
the general considerations about the chances of redeeming one's investment
remain true for all the contracts in our collection.
In a contract which poses no limits on how much one can invest, an
unusually high investment could make the contract stop sending payouts
for a lot of time, while accumulating the payout, thus discouraging new users to join.
Also, higher owner fees and higher multiplying factors will slow the flow down 
so that our results constitutes a lower bound to the number of users to be waited. 
For instance, the chain-shaped scheme \code{Doubler2} doubles the invested amount, 
asks a minimum investment of $1\ether$, and $10\%$ fee.  
The contract has paid out only up to the 68th user out of 210.  
Looking at its transactions, we see that the most common toll is between $1$ to $5\ether$, 
but here and there, there are some higher ones (up to $50$\ether) which make the system very slow to
fill up a level.

\paragraph{Tree-shaped  schemes}
The considerations above hold as a lower bound also for tree-shaped schemes, 
since they are slowed up by the fact that new users could not all be descendant of a given  node.

\paragraph{Waterfall schemes}
Assume a waterfall scheme with a fixed toll of $1$\ether, no fees, 
and which gives each user $10\%$ of the amount invested.  
For each new investor, the old ones are entitled of $0.1$\ether:
hence, $10$ new users are needed to repay the investment of the first user,
and further $10$ users to let her double her investment.
Note that for the first $10$ users, the amount they are giving is
not entirely distributed: a part is left within the contract.
However, after the $10$th investor joins the chain, the money she is
giving is not enough to be shared among {\it all} the users: from that
moment on, the contract must use its own funds to fill up the gap. 
Eventually, also this amount will end: as the scheme goes on, no
matter how many other investors will join, only the first $10$ users
are guaranteed to receive shares.

So, to ensure that {\it each} user can double her investment,
we must make sure that investments are spread over all the users. 
Let us now assume that, to join the scheme, a user must give
$0.1$\ether times number of users already in. 
So the first user will invest $0$, and the $11th$ user will invest $1$\ether.  
With such a rule, if the scheme contains $n$ users, the $k$-th user has given 
$0.1 (k-1)$\ether while receiving $0.1 (n - k)$\ether. 
As the number of users grows, also the received money grows. 
For instance, the 3rd user joining the scheme must invest $0.2$\ether,
and will receive $0.4$\ether as soon as $4$ other users join in.  
Instead, the 50th user must give a toll of $4.9$\ether, 
and must wait $98$ new users to double the investment.
However, since the toll increases for each new investor, 
entering the scheme is less appealing as the scheme goes on.

\paragraph{Handover  schemes}
In handover schemes, for an investor to receive a payout 
it is enough to wait one other user to join.  
However, since the toll keeps increasing, 
entering the scheme is less appealing as the scheme goes on.

\medskip
Overall, we have shown that the requirements \textbf{R3} and \textbf{R4} hold
for all the kinds of schemes we have identified:
namely, each investor is guaranteed to earn money if enough money are invested afterwards,
but late investors have a greater the risk of losing their investment.

\begin{figure}[t]
\centering
\iftoggle{forest}{
\begin{forest}
  colour me/.style={
    fill=#1,
    draw=#1,
  },
  for tree={
    circle,
    circular drop shadow,
    text=white,
    math content,
    edge+={draw=gray, thick},  
  },
  where n children=0{% for the leaves
    colour me=red,
    minimum size=2mm,
  }  {
    if level=0{% for the root
      colour me=orange,
      minimum size=18mm,
    }{
    if level=1{% 
      colour me=orange,
      minimum size=12mm,
      }{  if level=2{% 
        colour me=orange,
        minimum size=10mm,
       }{ 
         colour me=orange,
         minimum size=6mm,
       }       
     }
    },
  },
  delay={
    where content={k}{% just the empty nodes
      colour me=blue,
      minimum size=6mm,
    }{},
  },
  [1 
    [2 
      [4  [... [] []] [... [] []] ][5  [... [] [] ] [... [] []]]    
    ]
    [3 
      [6  [ k [] []][ ] ] [7  [] [ ] ]
    ]
  ]
\end{forest}
}
{}
\caption{Payout tree for a scheme which doubles the invested money and
  accepts exactly 1\ether from each user. 
  The first ether is given to the owner.}
\label{fig:ponziTree}
\end{figure}

\subsection{Security issues}
\label{sec:source-security}

In this~\namecref{sec:source-security} we highlight several
vulnerabilities we have encountered in the implementations of Ponzi schemes, 
which undermine their advertised behaviour. 
We will organize vulnerabilities according to their effects:
those harming investors, and those harming the scheme itself.

\paragraph{Harming  investors.}
Some vulnerabilities are due to bugs of the code, which is some cases seem
plain intentional: they harm users while being profitable for the owner of the scheme.
The most common vulnerability is caused by an improper use of the \code{send} primitive,
whose security issues have been already pointed out in~\cite{ABC17post,Luu16ccs}.  
If a \code{send} fails, it returns an error code: 
if a contract does not check this error, it cannot
acknowledge that there has been a problem.  So, in case of errors
during the \code{send}, the money remains within the contract, while
the user does not receive anything.  Notably, the large majority of
the contracts we have analysed do not check that the ether transfer succeeds.  
Their code is similar to that in \Cref{ex:tree-shaped}
(line~\lineno{27}), \Cref{ex:chain-shaped} (line~\lineno{22}),
\Cref{ex:waterfall} (line~\lineno{29}) and \Cref{ex:handover} (line~\lineno{13}).  
This vulnerability is known, at least, since February 11st 2016, 
when the owner of \code{KingOfTheEtherThrone} realized that there was too much ether left on his contract%
\footnote{Source:
  \href{https://www.reddit.com/r/ethereum/comments/44h1m1/}{\code{www.reddit.com/r/ethereum/comments/44h1m1/}}}.

Another issue affects many of the contracts which require a lower bound on the entry toll.
If the toll is not met, the user is not allowed to join the scheme, 
and the sent amount should be returned.
However, some contracts 
(\eg, \code{DynamicPyramid}, \code{GreedPit}, \code{NanoPyramid}, \code{Tomeka}), 
forget to retun it to user, and  keep the amount by themselves 
(see \eg~\Cref{ex:changing-percentage} left).
This is a questionable, especially when the minimum amount
is quite relevant (\eg, in \code{Tomeka} the minimum is 1 \ether).

Another bug that benefits the owner is in \code{PiggyBank}% 
\footnote{Source:
  \href{https://www.reddit.com/r/ethereum/comments/4br0za/piggybank_earn_eth_forever/}%
  {\code{www.reddit/piggybank\_earn\_eth\_forever}}.}.
According to its advertisement%
\footnote{Source: 
  \href{https://bitcointalk.org/index.php?topic=1410587.0}{\code{bitcointalk.org/topic=1410587.0}}.},
this is a waterfall scheme, where the owner keeps $3\%$ fees,
and each user receives $3\%$ of their investment every time 
a new user joins the scheme.
Hence, the command to compute the owner fees should be as follows:%

\begin{center}
\code{ fees = amount / 33}
\end{center}

\noindent
However, the actual command used in \code{PiggyBank} is just a little different:% 

\begin{center}
\code{fees += amount / 33}
\end{center}

\noindent
This difference is subtle to spot, but relevant:
indeed, the second command makes the fees grow at each deposit, 
and consequently the owner share subtracted to each investment steadily increases.
In practice, the fees calculated for the 7th deposit have exceeded the deposit itself.
Beside this, the global variable used to scan the array 
(\code{pos} in~\Cref{ex:waterfall}) is not reset, unlike in line~\lineno{25}.  
Hence, at each deposit, the iteration does not go from the \emph{first} user to the last one, 
but from the \emph{last} to the last itself.  
Hence, only one user at each deposit is paid, and only once.
Notably, the conjunction of these two bugs results in giving 
(almost) all the money invested to the owner. 
Were only the second bug present, the contract would have
kept accumulating a lot of unredeemable ether. 

Besides bugs hidden in the code, other dangers for users come from 
functions which allow the owner to perform special operations,
which can make the contract stray from its expected behaviour. 
One example is in \code{DynamicPyramid}, 
where the owner can change the interest rate, and also his fee shares
(see~\Cref{ex:changing-percentage}, right).
Other cases are in \code{Doubler3} and \code{TheGame}, where
the owner can withdraw all the money in the contract
(not only his share, see~\Cref{ex:termination}, left), 
draining the amount to be given back to investors.
Further, some schemes have a \code{selfdestruct} function that can be called only
by the owner, and terminates the contract (see~\Cref{ex:termination}, right). 
When this happens, investors lose their money.

\begin{figure}
  % \noindent\;
  \begin{minipage}[t]{0.5\textwidth}
\begin{lstlisting}[language=solidity,numbers=left,numbersep=10pt]
function init() private{
  //Ensures only tx with 1 ether
  if (msg.value < 1 ether) {
    collectedFees += msg.value;
    return;
  }...
\end{lstlisting}
  \end{minipage}
  \begin{minipage}[t]{0.6\textwidth}
\begin{lstlisting}[language=solidity,numbers=left,numbersep=10pt]
function changeMultiplier(uint _mult){
  if (msg.sender != owner) throw;
  if (_mult > 300 || _mult < 120) throw;
    pyramidMultiplier = _mult;
}
    
function changeFeePercentage(uint _fee){
  if (msg.sender != owner) throw;
  if (_fee > 10) throw;
  feePercent = _fee;
}
\end{lstlisting}
  \end{minipage}    
  \vspace{-15pt}
  \caption{On the left, rejecting enrollment without returning the fee in \code{Tomeka}. 
    On the right, the function used by the owner of \code{TheGame} to set multipliers and fees.}
  \label{ex:changing-percentage}
\end{figure}

\begin{figure}
  % \noindent\;
  \begin{minipage}[t]{0.45\textwidth}
\begin{lstlisting}[language=solidity,numbers=left,numbersep=10pt]
function Emergency() {
  if (owner!=msg.sender) throw;
  if (balance!=0){
    owner.send(balance);
    balance=0;
  }
}
\end{lstlisting}
  \end{minipage}
  \hspace{10pt}
  \begin{minipage}[t]{0.6\textwidth}
\begin{lstlisting}[language=solidity,numbers=left,numbersep=10pt]
function restart() {
  if (msg.sender==mainPlayer) {
    mainPlayer.send(address(this).balance);
    selfdestruct(mainPlayer);
  }
}
\end{lstlisting}
  \end{minipage}    
  \vspace{-15pt}
  \caption{On the left, withdrawing all the balance in \code{EthVentures1}.
    On the right, a termination function in \code{TheGame}.}
  \label{ex:termination}
\end{figure}

\begin{figure}[t]
  \begin{minipage}[t]{0.6\textwidth}
\begin{lstlisting}[language=solidity,numbers=left,numbersep=10pt]
contract HYIP {
uint constant INTERVAL = 1 days;

struct Investor {	
  address addr;
  uint amount;
}
Investor[] private investors;
address private owner;
uint private paidTime;

function HYIP() {
  owner = msg.sender;
  paidTime = now;		
}

function() payable {
  investors.push(Investor(msg.sender, msg.value) );
}

function performPayouts() {
  if(paidTime + INTERVAL > now)  throw;
    						
  uint fees = (this.balance * 37)/1000;
  if (!owner.send(fees)) throw;

  uint idx;
  for (idx = investors.length; idx-- > 0; ) {
    uint payout =
      (investors[idx].amount * 33) / 1000;
    if(!investors[idx].addr.send(payout))
      throw;		
  }			
  paidTime += INTERVAL;	
}}
\end{lstlisting}
  \end{minipage}
  \hspace{10pt}
  \begin{minipage}[t]{0.4\textwidth}
\begin{lstlisting}[language=solidity,numbers=left,numbersep=10pt]
contract Mallory {

  address victim = 0x23...;
  address private owner;
  bool private attack = true;

  //to be created with
  //1wei of balance
  function Mallory() {
    owner = msg.sender;
  }

  function() payable {
    if (attack) throw;
  }

  function invest() {
    victim.send(1 wei);
  }

  function stopAttack(){
    if (msg.sender == owner)
      attack = false;
  }
}
\end{lstlisting}
  \end{minipage}     
  \vspace{-5pt}
  \caption{On the left, a snippet of the code of \code{HYIP}, a scheme  vulnerable to Denial-of-Service attacks. On the right, the corresponding attack.}
  \label{ex:dos}
\end{figure}

\paragraph{Harming the scheme.}
Even when \code{send} commands are checked,
an improper handling of their return value can backfire, and can expose 
the scheme to Denial-of-Service attacks or blackmailing. 
An example is \code{HYIP} (see~\Cref{ex:dos}),
a waterfall scheme where investors are recorded in an array, 
and they are all paid at the end of every day. 
The scheme checks that each \code{send} is successful: 
in case of errors, it throws an exception.  
However, any error in one of the \code{send} (lines~\lineno{25} and~\lineno{31}) 
will revert \emph{all} the ether transfers.  
Errors may happen, \eg, for the following reasons:
\begin{inlinelist}
\item the array of investors grows so long that scanning it causes an out-of-gas exception; 
\item the balance of the contract goes to zero somehow in
  the middle of the \code{for} command (line~\lineno{28}), 
  having not paid all the investors; 
\item one of the investor is a contract, whose fallback raises an exception.  %
\end{inlinelist}
By exploiting the last issue, an attacker could create a contract
with a fallback which always \code{throw}s 
(see \eg, \code{Mallory} in~\Cref{ex:dos}). 
The attacker contract sends a fraction of ether to
\code{HYIP} to enter in the array of investors;
when \code{HYIP} tries to send her the payout, 
the invoked fallback throws an exception.  
Note that there is no way to cancel \code{Mallory} from the array, 
hence \code{HYIP} is stuck, and its balance is frozen forever. 
At this point, the attacker could blackmail \code{HYIP}, 
asking for money to stop the attack 
(via \code{stopAttack}, line~\lineno{21}).

Although the unchecked \code{send} is the most widespread issue,
there are other bugs which affect contracts. 
For instance, \code{Government}%
\footnote{\code{Government} is often called ``GovernMental'' or ``PonziGovernMental'' on web forums.}, 
has a notorious bug, which has been found, so far, only in that contract.
\code{Government} is a chain-shaped Ponzi scheme with a quirk: 
in addition to the usual way to get back money if enough
users keep investing, someone can win a jackpot if no one invests
after him for 12 hours. 
The list of users is kept in an array, and when the 12 hours have expired, 
the array is cleared. 
However, the command used to clear the array had to scan
\emph{each} of its elements. 
At a certain point, the array grew so long 
that clearing every element required too much gas ---
more than the maximum allowed per single transaction.  
Hence, the contract got stuck, with the legit jackpot winner unable to claim her price.

Another bug concerns the constructor function, % has the wrong name.
which is executed just once at creation time 
(usually, to initialize the owner of the contract with the address
\code{msg.sender} of the sender of the first transaction).  
The constructor must have the same name of the contract, 
but we found four contracts where it has a wrong name: 
\code{GoodFellas}, \code{Rubixi}, \code{FirePonzi}, and \code{StackyGame}.  
\Cref{ex:constructor-bug} shows an extract from the first two. On the
left, \code{Goodfellas} has a function called \code{LittleCactus} (line~\lineno{5}) 
which sets the owner, and then the owner is sent the fees collected so far
(line~\lineno{11}).  
On the right, \code{Rubixi} has a function called \code{DynamicPyramid} (line~\lineno{5}) 
which sets the owner (called \code{creator}), and then there is a function
\code{collectAllFees} which can be invoked to send the fees to the
owner (line~\lineno{11}). 
Giving a wrong name to the function meant to be a constructor is harmful: 
the function does not qualify to be a contructor at all, 
and it can be invoked by anyone at anytime, hence changing the owner address. 
When users discovered the bug, they started to invoke these functions to obtain
the ownership and redeem the fees.

To conclude this list of issues, we illustrate a simple trick that  can be performed
to shut down a  chain-shaped scheme.
To illustrate it, we consider \code{Doubler}, 
which sends back the amount multiplied by two.  
To perform the attack, Oscar needs to invest a large amount of ether 
(say, 100\ether). 
Oscar first sends 100\ether to the contract, and then additional
100\ether (plus some fees)%
\footnote{To guarantee the atomicity of the
  sends, Oscar sends the money through a contract.}. 
Upon receiving the second slot, the scheme will pay all the 
200\ether back to Oscar, so he does not lose anything. 
From that moment on, all the subsequent investments will be gathered 
to pay back the second 100\ether of Oscar. 
If the average invested amount is smaller than 100\ether, 
a large number of investors (and a lot of time) are needed to pay back Oscar: 
hence, the scheme will not be able to pay out other investors for a while. 
Since the success of these schemes is based on the fact that they
are fast to pay out, it is likely that with this attack, 
the scheme will be abandoned.
This attack can be performed at any time
to disincentivize users to join a chain-shaped scheme%
\footnote{As far as we know, this attack has been performed only 
  on contract \code{Quadrupler}. 
  See
  \href{https://etherscan.io/address/0xa379bbdd0af814502eb9b38d475c7fa7411bb4ec}{etherscan}  and
  \href{https://bitcointalk.org/index.php?topic=1426329.0}{bitcointalk} for details.}.
If performed at an early stage of the lifecycle of the scheme, 
the attack succeeds with a negligible loss of money.

\begin{figure}[t]
  \begin{minipage}[t]{0.5\textwidth}
\begin{lstlisting}[language=solidity,numbers=left,numbersep=10pt]
contract Goodfellas {

  address public owner;

  function LittleCactus() {
    owner = msg.sender;
  }

  function enter() {
    ...
    owner.send(collectedFees);
    ...
  }

\end{lstlisting}
  \end{minipage}
  \hspace{10pt}
  \begin{minipage}[t]{0.5\textwidth}
\begin{lstlisting}[language=solidity,numbers=left,numbersep=10pt]
contract Rubixi {
  address private creator;
       
  //Sets creator
  function DynamicPyramid() {
    creator = msg.sender;
  }
  //Fee functions for creator
  function collectAllFees()  {
    if (collectedFees == 0) throw;
    creator.send(collectedFees);
    collectedFees = 0;
  }
\end{lstlisting}
  \end{minipage}    
  \vspace{-5pt}
  \caption{Constructor bug in \code{Goodfellas} and \code{Rubixi}.}
  \label{ex:constructor-bug}
\end{figure}

\section{Impact of Ponzi schemes}
\label{sec:impact}

\begin{table}[t!]
\centering				
\small
\caption{Top-10 Ponzi schemes by amount of invested ether.}
\resizebox{\textwidth}{!}{
\begin{tabular}{lSSSSSSSSSS}
\toprule
& \multicolumn{2}{c}{{\#}Trans.} 
& \multicolumn{2}{c}{\ether}
& \multicolumn{2}{c}{\USD}
& \multicolumn{2}{c}{Users}
& \multicolumn{2}{c}{Transactions}
\\
\cmidrule(r){2-3}
\cmidrule(r){4-5}
\cmidrule(r){6-7}
\cmidrule(r){8-9}
\cmidrule(r){10-11}
{Contract name} & {in}   & {out}   & {in}   & {out}   & {in}     & {out}    & {paying}    & {paid} &  {first} & {last} \\ 
\midrule
\code{DynamicPyramid} 
& {444} & {143} & {7474} & {7437} & {84187} & {83541} & {175} & {51}
& {2016-02-23} & {2018-10-01}
\\
\code{DianaEthereum-x1.8} 
& {288} & {168}	& {5307} & {5303} & {61166} & {61266} & {129} & {84}
& {2016-03-08} & {2018-05-17}
\\
\code{Doubler2}	
& {395} & {161} & {4858} & {4825} & {26376} & {26220} & {211} & {68}
& {2016-02-16} & {2018-11-22}
\\
\code{ZeroPonzi}
& {627} & {499} & {4490} & {4489} & {49816} & {49770} & {47} & {28}
& {2016-04-04} & {2017-10-27}
\\
\code{Doubler}
& {156}	& {57} & {3073} & {3073} & {31292} & {35927} & {92} & {17}
& {2016-02-19} & {2018-06-26}
\\
\code{Government}
& {723} & {846} & {2939} & {2939} 
& {35738} & {40066} & {40} & {27}
& {2016-03-08} & {2017-03-20}
\\
\code{Rubixi}
& {686}	& {66} & {1367} & {1363} & {16986} & {16775} & {104} & {28}
& {2016-03-14} & {2019-01-24}
\\
\code{ProtectTheCastle2}
& {890} & {1257} & {1332} & {1332} & {186040} & {190802} & {101} & {68}
& {2016-03-20} & {2018-02-22}
\\
\code{EthereumPyramid}
& {978} & {339} & {986} & {917}	& {5044} & {5290} & {327} & {125}
& {2015-09-07} & {2018-04-11}
% \\
% \code{GreedPit}
% & {358}	& {66} & {918} & {895} & {10418} & {10194} & {61} & {13}
% & {2016-03-27} & {2016-10-14}
\\
\midrule
{\sc Total} (\totPonzi schemes)
& {\totInTx} & {\totOutTx} & {\totInETH} & {\totOutETH} & {\totInUSD} & {\totOutUSD} & {\totInUsr} & {\totOutUsr}
& {\text{---}} & {\text{---}}
\\
\bottomrule
\end{tabular}
} % resizebox
\label{fig:ponzi-stats}
\end{table}

In~\Cref{fig:ponzi-stats} we draw some general statistics about 
all the \totPonzi contracts in our collection,
and we give details about the first 10 contracts in our list,
ordered by total amount of invested ether. 
Full data about the collected Ponzi schemes,
including their unique addresses,
are reported online at~\urldataset.

The columns in~\Cref{fig:ponzi-stats} contain
the number of incoming and outcoming transactions,
and the overall transferred value, 
both in \ether and in \USD (rounded to an integer).
To convert the amount of each transaction to \USD, 
we use the average exchange rate on the day of the transaction,
obtained from~\href{https://etherscan.io}{\code{etherscan.io}}%
\footnote{Source: \url{https://etherscan.io/chart/etherprice}}.
The value transferred through a transaction
has a different meaning, according to whether the transaction is external or internal:
\begin{itemize}

\item \emph{external} transactions are created by users to invoke contract functions.
These transactions can transfer some ether from a user to the called contract.
Hence, this amount of ether is part of the \emph{inflow} of the contract (\ie, its incoming ether).

\item \emph{internal} transactions are of two kinds: the ones sent
  \emph{to} the contract under observation, and the ones sent \emph{from} it. 
  The first case happens when, instead of sending her money
  directly to a Ponzi scheme, a user goes through another contract
  (typically, a wallet contract): hence the amount linked to the
  transaction is part of the \emph{inflow} of the contract.
  The second case happens when the contract sends a payout to
  some user: in this case, the transaction amount is part of the
  \emph{outflow} of the contract.

\end{itemize}

Note that, similarly to~\cite{Vasek14fc} for Ponzi schemes on Bitcoin,
also in Ethereum we cannot precisely quantify the profit of scammers.
Indeed, it is not clear how to define \emph{who} is the scammer: 
of course the contract owner can be considered the originator of the scam, 
but he may have more than one addresses through which redeeming money.
Hence, we do not know how to separate % , in internal transactions, 
the money sent to legit users from the money sent to scammers.
A rough over-approximation of the profit of scammers 
is the total inflow of the scheme.

The columns ``Paying users'' and ``Paid users'' in~\Cref{fig:ponzi-stats}
indicate, respectively, the number of users who entered  the scheme 
(\ie, the distinct addresses that send money to the contract),
and the number of users that have received a payment from the contract.

The columns ``\USD'' and ``\ether'' in~\Cref{fig:ponzi-stats} 
give a first measure of the economic impact of Ponzi schemes on Ethereum.
Notice that \ether alone is not significant as a unit of measure:
actually, the exchange rate from \ether to \USD has been highly volatile,
as shown by the diagram in~\Cref{fig:eth-usd}.
Overall, we observe that the Ponzi schemes in our list
collected \mbox{\totInUSD\USD} from \mbox{\totInUsr} distinct users.
While the difference between incoming and outgoing \ether is always non-negative
(as contracts cannot send more \ether than what they receive),
the difference between incoming and outgoing \USD can be negative.
This is not a contradiction: it can be explained by the fact that 
the exchange rate between \ether and \USD has varied over time,
as depicted in~\Cref{fig:eth-usd}.

\begin{figure}[ht]
\centering
\scriptsize
\begin{tikzpicture}
\begin{axis}[
width  = 1\linewidth,
height = 5cm,
date coordinates in=x,
xtick=data,
xmajorgrids = true,
ymajorgrids = true,
xlabel absolute, xlabel style={yshift=-0.5cm},
ylabel absolute, ylabel style={yshift=0.2cm},
ymin=0,ymax=1100,ytick={100,200,...,1100},
xticklabel=\tiny\month.\year,
xticklabel style = {rotate=60,anchor=east},
enlargelimits = false,
xlabel={Date},
%y tick label style={/pgf/number format/1000 sep=},
scaled y ticks = false,
extra y tick style={grid=major, tick label style={xshift=-1cm}},
ylabel={Value (USD)},
]
\pgfplotstableread[col sep=comma]{results/eth-usd.csv}\data
\addplot [color=orange!100!white, thick, mark=none] table[x index = {0}, y index = {1}]{\data};
\end{axis}
\end{tikzpicture}
\vspace*{-5mm}
\caption{Ether/USD exchange rate (monthly average).}
\label{fig:eth-usd}
\end{figure}

\begin{table}[t!]
\centering				
\small
\caption{Statistics by kind of scheme.}
\resizebox{\textwidth}{!}{
\begin{tabular}{lSSSSSSSSS}
\toprule
& \multicolumn{1}{c}{{\#}Num} 
& \multicolumn{2}{c}{\ether}
& \multicolumn{2}{c}{\USD}
& \multicolumn{3}{c}{Users}
\\
\cmidrule(r){3-4}
\cmidrule(r){5-6}
\cmidrule(r){7-9}
{Kind} & {}   & {in}   & {out}   & {in}     & {out}    & {paying}    & {paid} & {\%}\\ 
\midrule
\textbf{Tree-shaped}
& {4} &  {410} & {400} & {1429} & {1416} & {161} & {83} & {51\%}
\\
\textbf{Chain-shaped} 
& {151} &  {41514} & {40170} & {587086} & {599347} & {1967} & {968} & {48\%}
\\
\textbf{Waterfall}	
& {4} &  {452} & {444} & {8836} & {11261} & {111} & {82} & {73\%}
\\
\textbf{Handover}	
& {4} &   {486} & {483} & {2618} & {3124} & {97} & {63} & {64\%}
\\
\textbf{Other}	
& {21} &  {1017} & {933} & {30693} & {87728} & {42} & {36} & {85\%}
\\
\midrule
{\sc Total} 
& {\totPonzi}  & {\totInETH}  & {\totOutETH} &  {\totInUSD} & {\totOutUSD} & {\totInUsr} & {\totOutUsr} & {51\%}
\\
\bottomrule
\end{tabular}
} % resizebox
\label{fig:statsByShape}
\end{table}

\subsection{ Statistics by schemes kind }

In~\Cref{fig:statsByShape} we display the statistics about
the impact of Ponzi schemes grouped by scheme kind, 
according to the taxonomy in~\Cref{sec:source}.
The columns show: the number of schemes for each kind;
the amount of incoming and outgoing \ether to the contract, 
and the corresponding values in USD; 
the number of users that invested in the scheme and that have received some payout; 
the {\it ratio} between these two values.

We see that, out of all \totPonzi contracts, almost the totality are
chain-shaped schemes ($151$). 
There are $12$ schemes between waterfall, tree-based and handover kinds,
while a portion of $21$ schemes has not found place in any of the analysed categories.  
Mostly they are experiments, or variants of existing schemes, or just
singularities in their own way.

The source code of Ponzi schemes falling  in the same category features only
slight differences between distinct instance: most contracts only
differ in the multiplication factor, in the applied fees, or in the
presence of auxiliary functions, like \eg fields getter/setter, 
or other utility functions for the owner.
The low variance in the code of Ponzi schemes is also witnessed
by the average normalized Levenshtein distance among their bytecode,
which is $0.54$, far less than \gldavg,
the average distance between the bytecode of two \emph{arbitrary} contracts.
This may suggest that, after the first Ponzi schemes have been created, 
the subsequent ones have been obtained by adapting the existing instances.

\section{Measuring gains and losses}
\label{sec:gains-losses}

\definecolor{col1}{RGB}{20, 20, 255}
\definecolor{col2}{RGB}{136, 136, 136}
\pgfplotsset{compat=newest}
\pgfplotsset{major grid style={gray!50}}

We now study the distribution of gains and losses among users.  
We expect to observe the common pyramidal pattern of Ponzi schemes, 
where only a few users make profit from their investment, 
while the vast majority loses money.

We start our analysis by considering two contracts:
\code{Doubler2} (a chain-shaped scheme) 
and \code{Etheramid1} (a tree-shaped scheme).

\Cref{fig:gain} shows the gain in ether for each
user who entered the scheme \code{Doubler2}. 
A negative amount indicates a loss.
The graph shows that the vast majority of users has a balance
close to zero, while just a few of them have a substantial gain, 
with a peak of $486 \ether$.
The integral of the graph is close to zero, 
\ie the contract has redistributed almost all the money it 
has received.

We now analyse the \emph{gain ratio},
\ie the ratio between received ether versus invested ether. 
According to advertisement, which promises to double the investment, 
each user should have a ratio of exactly $2$. 
Instead, \Cref{fig:ratio} shows that, out of a total of $210$ users,
$142$ never received any money back (the ones with label 0); 
$23$ have a gain ratio between $0$ and $1$ 
(meaning that they barely were able to regain what they had invested);
$44$ have a ratio between $1$ and $2$,
and only one has a very high ratio ($486$).

\begin{figure}
\hspace{-10pt}
\scriptsize
\centering
\begin{tikzpicture}
\begin{axis}[
    width  = 1\linewidth, 
    ymin=-320,
    ymax=520, 
    xtick=data, xticklabels = \empty,
    height = 4cm,  
    xmajorgrids = true, ymajorgrids = true,
    xlabel absolute, xlabel style={yshift=-0.5cm},
    y tick label style={font=\tiny,major tick length=0pt},
    enlargelimits = false,   scaled y ticks = false,
    extra y tick style={grid=major, tick label style={xshift=-1cm}},
    legend style={ ultra thin}, legend pos=north west,  
  ]
  \legend{\code{Doubler2}}
  \pgfplotstableread[col sep=comma]{results/gainLoss__Doubler2.csv}\data
  \addplot+[only marks, mark size=0.5, draw=col1] table[x index = {0}, y index = {1}] \data; 
\end{axis}
\end{tikzpicture}
\begin{tikzpicture}
\begin{axis}[
    width  = 1\linewidth, 
    ymin=-2,
    ymax=32,
    xtick=data, xticklabels = \empty,
    height = 4cm,  
    xmajorgrids = true, ymajorgrids = true,
    xlabel absolute, xlabel style={yshift=-0.5cm},
    y tick label style={font=\tiny,major tick length=0pt},
    enlargelimits = false,   scaled y ticks = false,
    extra y tick style={grid=major, tick label style={xshift=-1cm}},
    legend style={ ultra thin}, legend pos=north west,  
  ]
  \legend{\code{Etheramid1}}
  \pgfplotstableread[col sep=comma]{results/gainLoss__Etheramid1.csv}\data
  \addplot+[only marks, mark size=0.5, draw=col1] table[x index = {0}, y index = {1}] \data; 
\end{axis}
\end{tikzpicture}
\caption{Gain in \ether ($y$-axis) per user ($x$-axis).}
\label{fig:gain}
\end{figure}

\begin{figure}[t]
\centering
\begin{tikzpicture}
  \begin{axis}[ 
    height = 5.5cm, 
    ybar, ymin=0,
    flexible xticklabels from table={results/gainRatio__Doubler2__main.csv}{label}{col sep=comma},
    xticklabel style={font=\tiny}, y tick label style={font=\tiny}, 
    xtick=data,
    legend style={ultra thin}, legend pos=north east,  
    nodes near coords, nodes near coords align={vertical}
    ]
    \legend{\code{Doubler2}}
    \pgfplotstableread[col sep=comma]{results/gainRatio__Doubler2__main.csv}\datatable
    \addplot table[x expr=\coordindex,y=value]{\datatable};
  \end{axis}
\end{tikzpicture}
\begin{tikzpicture}
  \begin{axis}[ 
    height = 5.5cm, 
    ybar, ymin=0,
    flexible xticklabels from table={results/gainRatio__Etheramid1__main.csv}{label}{col sep=comma},
    xticklabel style={font=\tiny}, y tick label style={font=\tiny}, 
    xtick=data,
    legend style={ultra thin}, legend pos=north east,  
    nodes near coords, nodes near coords align={vertical}
    ]
    \legend{\code{Etheramid1}}
    \pgfplotstableread[col sep=comma]{results/gainRatio__Etheramid1__main.csv}\datatable
    \addplot table[x expr=\coordindex,y=value]{\datatable};
  \end{axis}
\end{tikzpicture}
\vspace{-10pt}
\caption{Number of users grouped by gain ratio for
  \code{Doubler2} and \code{Etheramid1}. 
  Label \emph{0} means no money has been received; 
  for $n>0$, label \emph{n} indicates a ratio between 
  $n-1$ (strict) and $n$ (included).  
  Label \emph{$\infty$} indicates users who
  have never sent anything but have received something.}
\label{fig:ratio}
\end{figure}

\begin{figure}[t]
\centering
\small
\begin{minipage}[t]{0.4\textwidth}
\begin{tabular}{|c|c|c||c|} 
 \hline 
 \multicolumn{4}{|c|}{  {\bf Doubler2} }\\
        Invested & Received& Net gain & Ratio\\
 \hline 
  1.0 &486.8 & 485.8& 486\\
  601 &1081.8 & 480.8& 1.2 \\ 
  370.6& 654.5 & 283.9&1.7 \\
  254 &457.2 & 203.2& 1.7\\
  200 &360& 159.9& 1.8\\
 \hline
\end{tabular}
\end{minipage}
\hspace{30pt}  
\begin{minipage}[t]{0.4\textwidth}
\begin{tabular}{ |c|c|c||c| } 
 \hline 
 \multicolumn{4}{|c|}{  {\bf Etheramid1} }\\
        Invested & Received& Net gain & Ratio\\
 \hline 
  0 & 30.6 & 30.6&  \emph{$\infty$}\\
  1 & 7 & 6& 7 \\ 
  1 & 6.4 & 5.4&6.4 \\
  1 & 6.2 & 5.2&6.2 \\
  2 & 6.6 & 4.6& 3.3\\
 \hline
\end{tabular}
\end{minipage}
\caption{Details of the $5$ users who have gained most
  for \code{Doubler2} and \code{Etheramid1}. 
  Each row represents one user. 
  Values are in \ether, cut to the first decimal non-zero digit.  
  Columns show: 
  (1) how much the user has sent to the contract; 
  (2) how much she has received, 
  (3) difference between columns 2 and 1; 
  (4) gain ratio.
  The ratio of users who have sent nothing is denoted by $\infty$.}
\label{fig:best-users}
\end{figure}

\Cref{fig:best-users} (left) shows the users with the highest gains. 
The user who earned $486 \ether$ has invested only $1 \ether$.
Subsequent users have a gain ratio strictly less than $2$: 
this means that to have such an high gain, 
they had to invest a lot of ether.
Since \code{Doubler2} is a chain-shaped scheme 
which doubles the investment,
from \Cref{fig:best-users} we infer that the gaining users
have invested more than once, 
and sometimes the contract has not given the promised ether back.

\Cref{fig:gain} shows the gain of each user in \code{Etheramid1}.
As before, the majority of users lost their money. 
Only a few users gained a little (up to $5 \ether$), 
and exactly one had an high income of $30 \ether$.
\Cref{fig:ratio} shows a behaviour similar to that of \code{Doubler2}.
From~\Cref{fig:best-users} we see that two users have a peculiar gain ratio: 
one received $7ETH$ upfront an investment of $1ETH$, 
and another one (probably, the initiator of the scheme) 
received $30.6ETH$ without investing money.

\medskip
We now consider the other contracts in our collection.
In~\Cref{app:charts} we show the diagrams of user gain,
gain ratio, and inflow-outflow, on a selection of $23$ contracts,
chosen among those with the most interesting features,
\eg high volume of payments, number of users, or number of transactions. 
The users-gains diagrams
show a similar pattern to \code{Doubler2}, 
notwithstanding the differences in number of users, scheme type and volume of ether
exchanged of the contracts we have selected. 
In general, 
the curve is quite shallow in the centre with a narrow positive peak on the far right,
and it stays mainly below the $x$-axis. 
This means that a lot of users lose money; 
a few gain something, and even less have extremely high incomes.
Also, the difference between inflow and outflow is almost zero 
--- if we include the owner among the users. 

From the gain ratio diagrams in~\Cref{app:charts},
we observe that the most numerous classes are
those of users who never received any money back, 
or have a ratio between $0$ and $1$. 
This means that the majority of users could not gain anything.  
The percentage of users not gaining anything is on average
around $70\%$ among the contracts in our selection. 
In particular, \Cref{fig:not-gaining} shows that the most unfair contract is
\code{Doubler} (where $88\%$ of users do not gain anything),
followed by \code{ShinySquirrels} ($87\%$) and \code{GreedPit} ($85\%$). 

From the joint analysis of the gain ratio diagrams and
of the users with highest gains (\Cref{fig:highest-gains}), 
we see that just one or two users per contract
have exceptionally high revenues,
and that, in some cases, 
they have not invested money. 
Generally, the exceptional high gains belongs to the owner, and
are due to the contract fees.
In some cases, there is more than one owner: 
\eg, \code{EthereumPyramid} has two owners, and \code{Rubixi} has six.  
However, the case of \code{Rubixi} is singular: a bug in the code allowed users
to steal the role of the owner and hence to receive the fees.
In other cases, the scheme has no owner fees, and hence the ratio graph
is more leveled (\eg, in \code{ZeroPonzi}, $59\%$ of users do not gain anything),
as the absence of fees allows for refunding more users.

\section{Evolution over time}
\label{sec:timing}

In this~\namecref{sec:timing} we study how Ponzi schemes behave over time. 
In particular, in~\Cref{sec:timing:in-out-correlation} we analyse the
timing correlation between inflow and outflow; 
in~\Cref{sec:timing:lifetime} we analyse the lifespan of Ponzi schemes; 
in~\Cref{sec:timing:volume-transactions} we observe 
how the volume of payments evolves over time.

\subsection{Inflow and outflow correlation}
\label{sec:timing:in-out-correlation}

We now study the behaviour of inflow and outflow transactions over time.
Similarly to the previous section, 
we first analyse the contracts \code{Doubler2} and \code{Etheramid1}, 
before discussing the general pattern.

\Cref{fig:flow} shows that \code{Doubler2} was active for $6$
months, with most of the activity concentrated in the first month.  
We see a correlation between the inflow and the outflow: each
outflow (red dot) is preceded by a sequence of inflows (blue dots) of
smaller amounts.
This is because chain-shaped schemes gather funds,
and then pay back a single user in a single shot.  Hence, the red dots
are higher in value, but less in number.
% While there is plenty of blue dots with no red one at the same moment;
% the converse does not apply: generally, no red dot is present which has not been
% triggered by a blue one. 
%
The red dots from 06-05 correspond to the contract owner
withdrawing the fees. 
\code{Etheramid} was active only for $1$ month, 
with 5 days of intense activity.
Also in this case there
is a strong correlation between inflow and outflow transactions:
however, the pattern is different from \code{Doubler2},
since new investments (blue spots) are immediately distributed among 
(some of) the users.  
This results in a set of red dots in decreasing value
whenever there is a blue dot.

\begin{figure}
\centering
\scriptsize
\resizebox{\linewidth}{!}{
\begin{tikzpicture}
\begin{axis}[ 
       width  = 1\linewidth, height = 3cm,
       date coordinates in=x, table/col sep=comma,
       xticklabel=\tiny \day-\month, 
       legend style={ ultra thin}, legend pos=north east,  
       xticklabel style={rotate=90, anchor=near xticklabel}]
   \legend{\code{Doubler2}}
   \pgfplotstableread[col sep=comma]{results/dateInFlow__Doubler2.csv}\data
   \addplot+[only marks,mark size=0.5, draw=blue] table[x index = {0}, y index = {1}] \data; 
   \pgfplotstableread[col sep=comma]{results/dateOutFlow__Doubler2.csv}\data
   \addplot+[only marks,mark size=0.5, draw=red] table[x index = {0}, y index = {1}] \data;
\end{axis}
\end{tikzpicture}}
\resizebox{\linewidth}{!}{
\begin{tikzpicture}
\begin{axis}[ 
       width  = 1\linewidth, height = 3cm,
       date coordinates in=x, table/col sep=comma,
       xticklabel=\tiny \day-\month, 
       legend style={ ultra thin}, legend pos=north east,  
       xticklabel style={rotate=90, anchor=near xticklabel}]
   \legend{\code{Etheramid1}}
   \pgfplotstableread[col sep=comma]{results/dateInFlow__Etheramid1.csv}\data
   \addplot+[only marks,mark size=0.5, draw=blue] table[x index = {0}, y index = {1}] \data; 
   \pgfplotstableread[col sep=comma]{results/dateOutFlow__Etheramid1.csv}\data
   \addplot+[only marks,mark size=0.5, draw=red] table[x index = {0}, y index = {1}] \data;
\end{axis}
\end{tikzpicture}}
\caption{Inflow (blue) and outflow (red) timing: 
on the $x$ axis, the time of transactions (day-month); 
on the $y$ axis, the amount of ether received/sent.}
\label{fig:flow}
\end{figure}

The diagrams of the other contracts in our selection are in~\Cref{app:charts}, 
from which we still see the strong correlation between inflow and outflow. 
The actual correlation depends on the category of the Ponzi scheme:
a tree-shaped one spreads each investment as soon as it receives
them (\eg, \code{Etheramid1}),
while a chain-shaped scheme waits to have enough balance.
Other minor differences in the diagrams depend on contract peculiarities. 
For instance, some contracts also implement a lottery,
so that there is a jackpot winner which results in an
unexpected red dot (\eg, \code{Government}). 
Other contracts are designed so that the owners can withdraw 
fees at their wish, and again in this
case, the result is red dots in a zone with no blue ones.
In general, most contracts have a very short lifespan, 
with a peak of intense activity followed by almost no activity at all.

\subsection{Lifetime}
\label{sec:timing:lifetime}

We now study the lifetime of the Ponzi schemes in our collection.
\Cref{fig:lifetime} displays, in blue, the lifetime measured as the
number of days from the first to the last inflow or outflow
transaction of the contract. 
We see that \mbox{$\sim 60\%$} of Ponzi schemes 
% ($87$ contracts out of \totPonzi)
have a lifetime close to 0 days.  
Basically, this means that they were deployed on the Ethereum
blockchain, and in many cases advertised in forums or dedicated web
sites, but they did not manage to attract any users.

Note that using the last transaction of a contract to
measure its actual period of activity may be too coarse.
Indeed, our overall diagrams show that Ponzi schemes 
are characterized by an high number of transactions 
operated in a short time frame,
followed by a period of isolated transactions, and inactivity.
% 
% The red curve of~\Cref{fig:lifetime} displays the lifespan of Ponzi
% schemes, trimmed to the first period of $10$ days of inactivity.  
% The result is a curve which better captures the duration 
% of the actual period of activity.

\begin{figure}[t]
  \centering
  \scriptsize
  \scalebox{1}{
    \begin{tikzpicture}
      \begin{axis}[
        width  = 1\linewidth,
        height = 6cm,
        legend pos= north east,
        ymin=0, ymax=1100,
        xmin=0, xmax=\value{valTotPonzi},
        ytick={0,200,400,..., 1100},
        xtick={0,20,40,60,80,100,120,140},
        enlarge y limits={abs=0.25cm},
        xlabel absolute, xlabel style={yshift=0.5cm},   
        ylabel=Days,
        xmajorgrids = true,
        ymajorgrids = true,
        ytick pos=left,
        legend entries = {\!\!Lifetime \;\;(avg = \avgPubLifetime)},
        % Trimmed lifetime\;\;(avg = \avgTrimmedLifetime)
        ]
        \pgfplotstableread[col sep=comma]{results/lifetime.csv}\data
        \addplot [blue,smooth,thick,mark=none] table[x index = {0}, y index = {4}]{\data};
        % 
        % \pgfplotstableread[col sep=comma]{results/lifetime-cropped.csv}\data
        % \addplot [red,dashed, smooth,thick,mark=none] table[x index = {0}, y index = {4}]{\data};
      \end{axis}
    \end{tikzpicture}
  }
  \caption{Lifetime of Ponzi schemes. 
    On the $x$-axis, the number of contracts; 
    on the $y$-axis, their lifetime measured in days.}
  \label{fig:lifetime}
\end{figure}

\Cref{fig:creation} shows how many Ponzi schemes have been created over time.
We see a peak in April 2016, with 91 new public Ponzi schemes.
After this first wave of creations, the situation has settled, 
with an average of $\sim 3$ new public schemes per month.
In~\Cref{sec:conclusions} we discuss possible explanations for this fall in the creation of 
Ponzi schemes, and in particular we conjecture that, rather than disappearing,
they are evolving into something more difficult to classify.

\begin{figure}[t]
  \centering
  \scriptsize
  \scalebox{1}{
    \begin{tikzpicture}
      \begin{axis}[
        width = 1\linewidth,
        height = 7cm,
        date coordinates in=x,
        xtick=data,
        ybar,
        bar width=5pt,
        % stack plots=y,
        xlabel absolute, xlabel style={yshift=-0.5cm},
        ylabel absolute, ylabel style={yshift=0cm},
        xticklabel=\tiny\month.\year,
        xticklabel style = {rotate=45,anchor=east},
        ytick={0,20,...,90},
        enlargelimits = true,
        % xmajorgrids = true,
        ymajorgrids = true,
        % title={Foo},
        % xlabel={Date},
        % y tick label style={/pgf/number format/1000 sep=},
        % scaled y ticks = false,
        % extra y tick style={grid=major, tick label style={xshift=-1cm}},
        % ylabel={Number of contracts},
        legend entries = {Number of Ponzi schemes},
        ]
        \pgfplotstableread[col sep=comma]{results/creation-public.csv}\data
        \addplot [fill=blue!50,draw=blue!50,solid,thick,mark=none] table[x index = {0}, y index = {1}]{\data};
      \end{axis}
    \end{tikzpicture}
  }
  \caption{Number of Ponzi schemes created by month.}
  \label{fig:creation}
\end{figure}

\subsection{Volume of payments}
\label{sec:timing:volume-transactions}

In this~\namecref{sec:timing:volume-transactions} we study how Ponzi
schemes perform over time.  \Cref{fig:ponzi:full-volume-by-time} shows
the daily volume of payments (measured in \USD) of all the \totPonzi
Ponzi schemes in our collection.
The $x$-axis represents time, and the $y$-axis gives the volume of
money transferred (measured in \USD).  The red dashed line represents
money sent by users to the schemes, while the blue solid line
represents money sent by the schemes to users.
The diagram clearly reports an equilibrium between outcoming and incoming flows, 
meaning that most of the money invested in the schemes are redistributed to users.
However, the distribution of money follows the pattern of inequality that characterizes 
Ponzi schemes, as highlighted in~\Cref{sec:gains-losses}, 
and further discussed later on in~\Cref{sec:inequality}.

From~\Cref{fig:ponzi:full-volume-by-time} we observe that most value
was exchanged in the period from February to May 2016,
with three peaks between March and April 2016.
It is plausible that the fall of activity after April 2016
is a consequence of the analogous drop in the creation of new Ponzi schemes,
witnessed in~\Cref{fig:creation}.

\begin{figure}[h]
\centering
\scriptsize
\begin{tikzpicture}
\begin{axis}[
date ZERO=2015-09-07,
xmin=2016-02-01,
xmax=2016-12-01,
xtick={2016-02-19,2016-03-11,2016-03-31,2016-04-04,2016-06-17},
width  = 1\linewidth,
height = 8cm,
date coordinates in=x,
%xmajorgrids = true,
%ymajorgrids = true,
%xticklabel style={rotate=90,anchor=near xticklabel},
xlabel absolute, xlabel style={yshift=-0.5cm},
ylabel absolute, ylabel style={yshift=0.2cm},
xticklabel=\tiny\year.\month.\day,
xticklabel style = {rotate=90,anchor=east},
% enlargelimits = false,
%title={Foo},
% xlabel={Date},
%y tick label style={/pgf/number format/1000 sep=},
scaled y ticks = false,
extra y tick style={grid=major, tick label style={xshift=-1cm}},
]
\pgfplotstableread[col sep=comma]{results/in__tot.csv}\data
\addplot [color=blue,thick] table[x index = {0}, y index = {1}]{\data};
\pgfplotstableread[col sep=comma]{results/out__tot.csv}\data
\addplot [color=red,thick,dashed] table[x index = {0}, y index = {1}]{\data};
\legend{Incoming \USD,Outcoming \USD}
\end{axis}
\end{tikzpicture}
\caption{Daily volume of transactions (complete set of \totPonzi Ponzi schemes).}
\label{fig:ponzi:full-volume-by-time}
\end{figure}

We now measure the volume of transactions pointwise, 
on a sample of the most representative schemes.  
Each diagram in~\Cref{fig:ponzi:volume-by-time} shows the money flow (in and out)
of a single contract: the red dashed lines represent money sent to the
scheme (measured in \USD), while the blue solid lines represent
payouts sent by the scheme to users.  
The $x$-axis represents time: we
consider the total incoming/outgoing money per day.

In the diagram for \code{DynamicPyramid}, we see that the most of the
incoming flow happened on the 11st of March, 
the total investments in that day alone amount to almost
$\USDfmt{60000}\USD$%
\footnote{Source: \href{http://etherscan.io/address/0xa9e4e3b1da2462752aea980698c335e70e9ab26c}{\code{etherscan.io/address/0xa9e4e3b1da2462752aea980698c335e70e9ab26c}}.}.
We see that the blue and red flows in that day almost overlap, 
meaning that with such a great balance the contract
was able to pay out many users. 
However, after that single peculiar day, users almost stopped sending ether, 
and so did the contract. 
% Now the contract has a balance of $37$\ether, and it is waiting to pay
% out the 52nd user of 174.

The diagram of \code{Government} is peculiar, 
due to a bug which affected it
(already discussed in~\Cref{sec:source-security}).
This contract needs to periodically clear the array which records the list of users.
However, from a certain point performing this operation would have required more gas than the
maximum allowed for a single transaction.
Several attempts to clear the array and to redeem the funds stored in the contract have failed 
with an ``out-of-gas'' exception. 
Exactly in the date of the first hard-fork (on the 17th of June, 2016),
which also raised the gas limit%
\footnote{Source: \href{https://blog.ethereum.org/2016/07/20/hard-fork-completed/}{\code{blog.ethereum.org/2016/07/20/hard-fork-completed/}}},
we observe an internal transaction of
$\USDfmt{22699}\USD$, used to withdraw the funds and correctly clear the array.

The diagram for \code{EthereumPyramid} shows that the investments
were made basically in two slots of time: one around the last days of
February 2016, and another on a single day, the 1st of April. 
All those later investments were strangely made by the owner: 
they were almost 50 in a single day. 
We see that the inflow and outflow almost overlap.
\code{EthereumPyramid} asks to all users exactly 1\ether and triples the investment. 
With such a fixed toll, one user every three is paid out, and the outflow is smooth.  
However, we see that there is a peak in the outflow around the 26th of June, 
and in that day we observe a single payment of $90$\ether. 
After inspecting the code and the set of transactions, 
we are inclined to say that it is the owner withdrawing her fees.

From the diagram of \code{Etheramid} we see a perfect overlap of inflow and outflow: 
indeed, this is a tree-shaped scheme, so everything which goes in is immediately sent to the
users' ancestors. 
There is no need to delay payments waiting for the payout to reach its quote, 
like in chain-shaped schemes (\eg, see the diagram of \code{Doubler2}).

\begin{figure}[htbp!]
\begin{tabular}{l}

\begin{minipage}{\linewidth}
\centering
\scriptsize
\begin{tikzpicture}
\pgfplotsset{title style={at={(0.5,0.7)}}}
\begin{axis} [
title = {\code{DynamicPyramid}},
% ybar,
% bar width=0pt,
date coordinates in=x,
xtick={},
date ZERO=2016-02-23,
xmin=2016-03-01,
xmax=2016-06-15,
xtick={2016-03-11,2016-04-10,2016-05-10},
width  = \linewidth,
height = 4.5cm,
legend pos=north west,
scaled y ticks = false,
legend columns=1,
xlabel absolute, xlabel style={yshift=+0.2cm},
ylabel absolute, ylabel style={yshift=-0.2cm},
mark size=1.2pt,
]
\pgfplotstableread[col sep=comma]{results/in__0xa9e4e3b1da2462752aea980698c335e70e9ab26c.csv}\data
\addplot [draw=blue,thick] table[x index = {0}, y index = {1}]{\data};
\pgfplotstableread[col sep=comma]{results/out__0xa9e4e3b1da2462752aea980698c335e70e9ab26c.csv}\data
\addplot [draw=red,thick,dashed] table[x index = {0}, y index = {1}]{\data};
\end{axis}
\end{tikzpicture}
\end{minipage}

\\
\\

\begin{minipage}{\linewidth}
\centering
\scriptsize
\begin{tikzpicture}
\pgfplotsset{title style={at={(0.5,0.7)}}}
\begin{axis} [
title = {\code{Government}},
date coordinates in=x,
date ZERO=2016-03-08,
xmin=2016-03-08,
xmax=2016-06-24,
xtick={2016-03-10,2016-06-17},
width  = \linewidth,
height = 4.5cm,
legend style={at={(0.5,1.03)},anchor=south},
scaled y ticks = false,
legend columns=2,
xlabel absolute, xlabel style={yshift=+0.2cm},
ylabel absolute, ylabel style={yshift=-0.2cm},
%yaxis line style = ultra thick,
%x tick label style={/pgf/number format/.cd, fixed, fixed zerofill,precision=0},
%xmin=-12,xmax=12,xtick={-11,-9,...,11},
%ymin=,ymax=1.2,,ytick={0.2,0.4,...,1.0},
% xmajorgrids = true,
% ymajorgrids = true,
mark size=1.2pt,
]
\pgfplotstableread[col sep=comma]{results/in__0xf45717552f12ef7cb65e95476f217ea008167ae3.csv}\data
\addplot [draw=blue,thick] table[x index = {0}, y index = {1}]{\data};
\pgfplotstableread[col sep=comma]{results/out__0xf45717552f12ef7cb65e95476f217ea008167ae3.csv}\data
\addplot [draw=red,thick,dashed] table[x index = {0}, y index = {1}]{\data};
% \legend{Payments out, Payments in}
\end{axis}
\end{tikzpicture}
\end{minipage}

\\
\\

\begin{minipage}{\linewidth}
\centering
\scriptsize
\begin{tikzpicture}
\pgfplotsset{title style={at={(0.5,0.7)}}}
\begin{axis} [
title = {\code{EthereumPyramid}},
date coordinates in=x,
xtick={},
date ZERO=2015-09-07,
xmin=2015-11-07,
xmax=2016-07-10,
xtick={2016-02-27,2016-04-01,2016-06-24},
width  = \linewidth,
height = 4.5cm,
legend pos=north west,
scaled y ticks = false,
legend columns=1,
xlabel absolute, xlabel style={yshift=+0.2cm},
ylabel absolute, ylabel style={yshift=-0.2cm},
%yaxis line style = ultra thick,
%x tick label style={/pgf/number format/.cd, fixed, fixed zerofill,precision=0},
%xmin=-12,xmax=12,xtick={-11,-9,...,11},
%ymin=,ymax=1.2,,ytick={0.2,0.4,...,1.0},
% xmajorgrids = true,
% ymajorgrids = true,
mark size=1.2pt,
]
\pgfplotstableread[col sep=comma]{results/in__0x7011f3edc7fa43c81440f9f43a6458174113b162.csv}\data
\addplot [draw=blue,thick] table[x index = {0}, y index = {1}]{\data};
\pgfplotstableread[col sep=comma]{results/out__0x7011f3edc7fa43c81440f9f43a6458174113b162.csv}\data
\addplot [draw=red,thick,dashed] table[x index = {0}, y index = {1}]{\data};
%\pgfplotstableread[col sep=comma]{results/avg__0x7011f3edc7fa43c81440f9f43a6458174113b162.csv}\data
% \legend{EthereumPyramid}
\end{axis}
\end{tikzpicture}
\end{minipage}

\\
\\

\begin{minipage}{\linewidth}
\centering
\scriptsize
\begin{tikzpicture}
\pgfplotsset{title style={at={(0.5,0.7)}}}
\begin{axis} [
title = {\code{Etheramid1}},
date coordinates in=x,
xtick={},
date ZERO=2016-04-15,
xmin=2016-04-15,
xmax=2016-05-11,
xtick={2016-04-21,2016-04-30},
width  = \linewidth,
height = 4.5cm,
legend pos=north west,
scaled y ticks = false,
legend columns=1,
xlabel absolute, xlabel style={yshift=+0.2cm},
ylabel absolute, ylabel style={yshift=-0.2cm},
%yaxis line style = ultra thick,
%x tick label style={/pgf/number format/.cd, fixed, fixed zerofill,precision=0},
%xmin=-12,xmax=12,xtick={-11,-9,...,11},
%ymin=,ymax=1.2,,ytick={0.2,0.4,...,1.0},
% xmajorgrids = true,
% ymajorgrids = true,
mark size=1.2pt,
]
\pgfplotstableread[col sep=comma]{results/in__0x9758da9b4d001ed2d0df46d25069edf53750767a.csv}\data
\addplot [draw=blue,thick] table[x index = {0}, y index = {1}]{\data};
\pgfplotstableread[col sep=comma]{results/out__0x9758da9b4d001ed2d0df46d25069edf53750767a.csv}\data
\addplot [draw=red,thick,dashed] table[x index = {0}, y index = {1}]{\data};
\end{axis}
\end{tikzpicture}
\end{minipage}

\\
\\

\begin{minipage}{\linewidth}
\centering
\scriptsize
\begin{tikzpicture}
\pgfplotsset{title style={at={(0.5,0.7)}}}
\begin{axis} [
title = {\code{Doubler2}},
date coordinates in=x,
xtick={},
date ZERO=2016-02-16,
xmin=2016-02-16,
xmax=2016-06-10,
xtick={2016-03-01,2016-04-01,2016-05-01,2016-06-01},
width  = \linewidth,
height = 4.5cm,
legend pos=north west,
scaled y ticks = false,
legend columns=1,
xlabel absolute, xlabel style={yshift=+0.2cm},
ylabel absolute, ylabel style={yshift=-0.2cm},
%yaxis line style = ultra thick,
%x tick label style={/pgf/number format/.cd, fixed, fixed zerofill,precision=0},
%xmin=-12,xmax=12,xtick={-11,-9,...,11},
%ymin=,ymax=1.2,,ytick={0.2,0.4,...,1.0},
% xmajorgrids = true,
% ymajorgrids = true,
mark size=1.2pt,
]
\pgfplotstableread[col sep=comma]{results/in__0xfd2487cc0e5dce97f08be1bc8ef1dce8d5988b4d.csv}\data
\addplot [draw=blue,thick] table[x index = {0}, y index = {1}]{\data};
\pgfplotstableread[col sep=comma]{results/out__0xfd2487cc0e5dce97f08be1bc8ef1dce8d5988b4d.csv}\data
\addplot [draw=red,thick,dashed] table[x index = {0}, y index = {1}]{\data};
\end{axis}
\end{tikzpicture}
\end{minipage}

\end{tabular}
 
\caption{Volume of payments into and out Ponzi schemes, by time. %
On the $x$-axis, the dates of transactions; on the $y$-axis, the \USD sent to (blue solid line) and from (red dashed line) the contract.}
\label{fig:ponzi:volume-by-time}
\end{figure}

\section{Measuring payment inequality}
\label{sec:inequality}

Our last analysis measures the inequality in the distribution 
of investments and revenues for the schemes in our sample.
To this purpose we use 
\emph{Lorenz curves} (\Cref{fig:lorenz:in,fig:lorenz:out}) and \emph{Gini coefficients} (\Cref{fig:gini}),
two standard graphical representations of the distribution of income or wealth.

The Lorenz curves represent users on the $x$-axis (in percentage),
and on the $y$-axis the percentage of payments \emph{into} (\Cref{fig:lorenz:in}) 
and \emph{from} (\Cref{fig:lorenz:out}) the Ponzi scheme.
A diagonal line at 45 degrees from the two extremes of the diagram 
(leftmost-bottomost to rightmost-topmost)
represents the perfect equality: 
\ie, for all $x \in [0,100]$, the $x\%$ of the whole population of users 
has invested/received the $x\%$ of the total income of the scheme.
Instead, the perfect disequality is represented by the (discontinuous) function that
has value $0$ for all $x<100$, and value $100$ for $x=100$:
this means that a single user has invested/received the total sum in the scheme.

We can observe in~\Cref{fig:lorenz:in} that 
\code{Etheramid1}
is quite close to perfect equality,
while the most unbalanced schemes in our sample are 
\code{Government} and \code{ProtectTheCastle},
where $10\%$ of victims have invested more than $90\%$ of the money.
The Lorenz curves of these two schemes are quite close to
the overall curve of Bitcoin-only Ponzi schemes in~\cite{Vasek15fc}.
Overall, the closer is a curve to the one which represents perfect inequality,
the more a Ponzi scheme benefits from ``big fishes'' who invest large amounts of money in the scheme;
dually, if the curve is close to the one which represents perfect equality,
the scheme benefits from a large population of victims who invest a small amount of money.

From~\Cref{fig:lorenz:out} we observe that the distribution of payouts
is in general more iniquitous than that of investments,
as the Lorenz curves are more squeezed to the right, 
compared to those in~\Cref{fig:lorenz:in}.
Interestingly enough, although \code{Etheramid1}
is almost perfectly balanced for investments,
the distribution of payouts is quite unbalanced.

The Gini coefficients in~\Cref{fig:gini} relate the inequality of investments/payouts 
to the ``success'' of the scheme, defined as total amount of money invested/received by users.
The \mbox{$x$-axis} represents the degree of inequality
(0 indicates perfect equality, while 100 is perfect inequality),
and the $y$-axis measures the total investment/payout.
Each scheme is represented by an arrow, 
whose tail represents investments, while the head represents payouts.
For the most lucrative scheme, \code{DynamicPyramid},
we observe that the index of inequality is high, surpassing $80\%$
for both investments and payouts.
For \code{ProtectTheCastle}, we see that the head and the tail of the arrow almost overlap,
meaning that the inequality distributions of investments and payouts
are very close in this scheme.
For the less lucrative schemes, no correlation seems to exist
bewteen the success of the scheme and the index of inequality.

\begin{figure}[htbp!]
\centering
\scriptsize
\begin{tikzpicture}
\begin{axis}[
width  = 0.825\linewidth,
legend columns=2, 
legend pos= north west,
every axis legend/.append style={nodes={right}},
xmin=0,xmax=100,xtick={10,20,...,100},
ymin=0,ymax=100,ytick={10,20,...,100},
xmajorgrids = true,
ymajorgrids = true,
xlabel={Fraction of users ($\%$)},
ylabel={Fraction of payments into schemes ($\%$)},
]
% DynamicPyramid
\pgfplotstableread[col sep=comma]{results/lorenz__0xa9e4e3b1da2462752aea980698c335e70e9ab26c.csv}\data
\addplot [color=red, thick, dotted, mark=none] table[x index = {0}, y index = {1}]{\data};
% Government
\pgfplotstableread[col sep=comma]{results/lorenz__0xf45717552f12ef7cb65e95476f217ea008167ae3.csv}\data
\addplot [color=blue, thick, mark=none] table[x index = {0}, y index = {1}]{\data};
% EthereumPyramid
\pgfplotstableread[col sep=comma]{results/lorenz__0x7011f3edc7fa43c81440f9f43a6458174113b162.csv}\data
\addplot [color=teal, thick, dashed, mark=none] table[x index = {0}, y index = {1}]{\data};
% ProtectTheCastle
\pgfplotstableread[col sep=comma]{results/lorenz__0x7d56485e026d5d3881f778e99969d2b1f90c50af.csv}\data
\addplot [color=magenta, thick, dashdotted, mark=none] table[x index = {0}, y index = {1}]{\data};
% Doubler2
\pgfplotstableread[col sep=comma]{results/lorenz__0xfd2487cc0e5dce97f08be1bc8ef1dce8d5988b4d.csv}\data
\addplot [color=cyan, thick, dashdotdotted, mark=none] table[x index = {0}, y index = {1}]{\data};
% Etheramid1
\pgfplotstableread[col sep=comma]{results/lorenz__0x9758da9b4d001ed2d0df46d25069edf53750767a.csv}\data
\addplot [color=gray, thick, densely dotted, mark=none] table[x index = {0}, y index = {1}]{\data};
\legend{
\code{DynamicPyramid}, \code{Government}, \code{EthereumPyramid}, \code{ProtectTheCastle}, \code{Doubler2}, \code{Etheramid1}
}
\end{axis}
\end{tikzpicture}
\vspace{-10pt}
\caption{Lorenz curves of a sample of Ponzi schemes (payments in).}
\label{fig:lorenz:in}

\bigskip

\centering
\scriptsize
\begin{tikzpicture}
\begin{axis}[
width  = 0.825\linewidth,
legend columns=2, 
legend pos= north west,
every axis legend/.append style={nodes={right}},
xmin=0,xmax=100,xtick={10,20,...,100},
ymin=0,ymax=100,ytick={10,20,...,100},
xmajorgrids = true,
ymajorgrids = true,
xlabel={Fraction of users ($\%$)},
ylabel={Fraction of payments from schemes ($\%$)},
]
% DynamicPyramid
\pgfplotstableread[col sep=comma]{results/lorenz-out__0xa9e4e3b1da2462752aea980698c335e70e9ab26c.csv}\data
\addplot [color=red, thick, dotted, mark=none] table[x index = {0}, y index = {1}]{\data};
% Government
\pgfplotstableread[col sep=comma]{results/lorenz-out__0xf45717552f12ef7cb65e95476f217ea008167ae3.csv}\data
\addplot [color=blue, thick, mark=none] table[x index = {0}, y index = {1}]{\data};
% EthereumPyramid
\pgfplotstableread[col sep=comma]{results/lorenz-out__0x7011f3edc7fa43c81440f9f43a6458174113b162.csv}\data
\addplot [color=teal, thick, dashed, mark=none] table[x index = {0}, y index = {1}]{\data};
% ProtectTheCastle
\pgfplotstableread[col sep=comma]{results/lorenz-out__0x7d56485e026d5d3881f778e99969d2b1f90c50af.csv}\data
\addplot [color=magenta, thick, dashdotted, mark=none] table[x index = {0}, y index = {1}]{\data};
% Doubler2
\pgfplotstableread[col sep=comma]{results/lorenz-out__0xfd2487cc0e5dce97f08be1bc8ef1dce8d5988b4d.csv}\data
\addplot [color=cyan, thick, dashdotdotted, mark=none] table[x index = {0}, y index = {1}]{\data};
% Etheramid1
\pgfplotstableread[col sep=comma]{results/lorenz-out__0x9758da9b4d001ed2d0df46d25069edf53750767a.csv}\data
\addplot [color=gray, thick, densely dotted, mark=none] table[x index = {0}, y index = {1}]{\data};
\legend{
\code{DynamicPyramid}, \code{Government}, \code{EthereumPyramid}, \code{ProtectTheCastle}, \code{Doubler2}, \code{Etheramid1}
}
\end{axis}
\end{tikzpicture}
\vspace{-10pt}
\caption{Lorenz curves of a sample of Ponzi schemes (payments out).}
\label{fig:lorenz:out}
\end{figure}

\begin{figure}[h]
\scriptsize
\begin{tikzpicture}
\begin{axis}[
legend columns=3, 
legend style={/tikz/column 2/.style={column sep=5pt,},},
legend style={at={(0.5,1.3)},anchor=north},
width  = 1\linewidth,
height = 10cm,
%legend pos= north west,
every axis legend/.append style={nodes={right}},
scaled y ticks = false,
%ymode=log,
%log basis y={10},
ylabel absolute, ylabel style={yshift=0.0cm},
x tick label style={/pgf/number format/.cd,fixed,fixed zerofill,precision=0,/tikz/.cd},
%y tick label style={/pgf/number format/.cd,fixed,fixed zerofill,precision=0,/tikz/.cd},
xmin=0,xmax=100,xtick={10,20,...,100},
ymin=-3000,ymax=95000,
xminorgrids = true,
yminorgrids = true,
xmajorgrids = false,
ymajorgrids = false,
%scatter/classes={a={mark=circle,draw=gray}},
xlabel={Gini coefficient (\%)},
ylabel={Payments from/to scheme in \USD},
nodes near coords,
]

% DynamicPyramid
\addplot[red,->,thick,fill=white,point meta=explicit symbolic]
coordinates {
(81.84,83830)
(89.38,83408)
}
[yshift=8pt]
node [pos=0.5] {\code{DynamicPyramid}}; 
% \addlegendentry{DynamicPyramid}

% Government
\addplot[blue,->,thick,fill=white,point meta=explicit symbolic]
coordinates {
(89.30,36089)
(91.80,45758)
}
[yshift=15pt]
node [pos=0.5] {\code{Government}};
; 
% \addlegendentry{Government}

% EthereumPyramid
\addplot[teal,->,thick,fill=white,point meta=explicit symbolic]
coordinates {
(72.15,4875)
(90.07,4048)
}
[yshift=8pt]
node [pos=0.5] {\code{EthereumPyramid}};
% \addlegendentry{EthereumPyramid}

% ProtectTheCastle
\addplot[magenta,only marks,mark=x,thick,fill=white,point meta=explicit symbolic]
coordinates {
(91.84,12002)
(92.01,12092)
}
[yshift=8pt]
node [pos=0] {\hspace{-20pt}\code{ProtectTheCastle}};
% \addlegendentry{ProtectTheCastle}

% Doubler2
\addplot[cyan,->,thick,fill=white,point meta=explicit symbolic]
coordinates {
(79.77,24253)
(93.25,24232)
}
[yshift=8pt]
node [pos=0.5] {\code{Doubler2}};
% \addlegendentry{Doubler2}

% Etheramid1
\addplot[gray,->,thick,fill=white,point meta=explicit symbolic]
coordinates {
(8.02,940)
(79.62,676)
}
[yshift=8pt]
node [pos=0.5] {\code{Etheramid1}};
% \addlegendentry{Etheramid1}

% DynamicPyramid
%\pgfplotstableread[col sep=comma]{results/gini__0xa9e4e3b1da2462752aea980698c335e70e9ab26c.csv}\data
%\addplot [only marks,mark=asterisk,red,thick,fill=white,point meta=explicit symbolic] table[x index = {0}, y index = {1}]{\data};
% Government
% \pgfplotstableread[col sep=comma]{results/gini__0xf45717552f12ef7cb65e95476f217ea008167ae3.csv}\data
% \addplot [only marks,mark=square*,blue,thick,fill=white] table[x index = {0}, y index = {1}]{\data};
% EthereumPyramid
% \pgfplotstableread[col sep=comma]{results/gini__0x7011f3edc7fa43c81440f9f43a6458174113b162.csv}\data
% \addplot [only marks,mark=triangle*,green,thick,fill=white] table[x index = {0}, y index = {1}]{\data};
% % ProtectTheCastle
% \pgfplotstableread[col sep=comma]{results/gini__0x7d56485e026d5d3881f778e99969d2b1f90c50af.csv}\data
% \addplot [only marks,mark=o,magenta,thick,fill=white] table[x index = {0}, y index = {1}]{\data};
% % Doubler2
% \pgfplotstableread[col sep=comma]{results/gini__0xfd2487cc0e5dce97f08be1bc8ef1dce8d5988b4d.csv}\data
% \addplot [only marks,mark=pentagon,cyan,thick, fill=white] table[x index = {0}, y index = {1}]{\data};
% % Etheramid1
% \pgfplotstableread[col sep=comma]{results/gini__0x9758da9b4d001ed2d0df46d25069edf53750767a.csv}\data
% \addplot [only marks,mark=x,gray, thick, fill=white] table[x index = {0}, y index = {1}]{\data};
% \legend{DynamicPyramid, Government}
%, EthereumPyramid, ProtectTheCastle, Doubler2, Etheramid1
\end{axis}
\end{tikzpicture}
\caption{Gini coefficients of a sample of Ponzi schemes.}
\label{fig:gini}
\end{figure}

\section{Conclusions}
\label{sec:conclusions}

Blockchains and smart contracts might really be the next ``disruptive'' technologies,
as often reported by the media;
however, they can also offer new opportunities
to tax-evaders, criminals, and fraudsters,
who can take advantage of their anonymity and decentralization~\cite{Brito2013bitcoin,Slattery2014taking}.
In this paper we have analysed Ponzi schemes on Ethereum, 
the most widespread platform for smart contracts so far.
Overall, we have observed that, in the first \ethlifetimeyears years of life of Ethereum,
there have been a multitude of experiments to implement Ponzi schemes as smart contracts.
% indeed, \fracPonziVerif of the \totVerified contracts with verified source code 
% on \href{https://etherscan.io/}{\code{etherscan.io}}
% are Ponzi schemes.
Although the economic impact of these experiments has been quite limited,
as they involve only a small fraction % \fracPonziTx 
of the transactions and value on the Ethereum blockchain,
our analysis allows to draw some relevant conclusions, 
which we summarise below in the form of ``recommendations'' for users and surveillance authorities.

\paragraph{Recommendation \#1: check the advertisements.}

During our collection activity, we have studied how Ponzi schemes are promoted on the web.
In many cases, Ponzi schemes are presented as ``high-yield'' investment programs,
promising high returns and omitting to declare any risks;
in some other cases, they are promoted as mere ``social games'',
but a constant factor is that playing involves transfers of money from the user to the contract,
and the allurement of making some profits.
In many cases, we have found discrepancies between the advertisement
and the actual chances of obtaining a payout:
the latter is presented as a plain fact, while in~\Cref{sec:source} 
we have shown that fallacies in the money distribution mechanism or in its implementation
might prevent users from obtaining the expected payouts.
Further, advertisements usually omit to declare that the contract owner
can modify the advertised conditions, \eg by increasing the owner fees,
or destructing the contract%
\footnote{For instance, the advertisement of
\href{https://www.reddit.com/r/ethtrader/comments/4ds0a5/ethstick_a_satirical_yet_profitable_ponzi_game/}{\code{EthStick}}
only warns that ``the settings can be changed to adapt to the trends (but only within defined limits)''.
Actually, from its source code we see that the fees can only be augmented, while the payoff multiplier factor
can be only decreased: everything to the sole advantage of the owner.}.

So, our first recommendation for potential users is to carefully study the advertisement:
if the conditions appear too alluring, probably it is a scam.
Websites like \emph{BadBitcoin}, which maintains a blacklist of cryptocurrency-based scams%
\footnote{\url{https://badbitcoin.org/thebadlist/}},
or discussion forums like the ``Gambling: Investor-based games'' section of \emph{Bitcointalk.org}%
\footnote{\url{https://bitcointalk.org/index.php?board=207.0}}
should be consulted before sending money to a contract.
For surveillance authorities, our recommendation is to monitor the web 
to detect suspect advertisements, and to provide the community with official blacklists.

% Scam accusations https://bitcointalk.org/index.php?board=83.0
% Gambling: Games and Rounds https://bitcointalk.org/index.php?board=71.0
% Gambling: Investment Games https://bitcointalk.org/index.php?board=207.0

\paragraph{Recommendation \#2: analyse the contract code.}
Despite one of the main selling points of ``smart'' Ponzi schemes
is that their immutability and decentralised execution makes them ``reliable'',
our analysis in~\Cref{sec:source} has revealed several vulnerabilities,
which undermine their trustworthiness.
Some of these vulnerabilities are caused by poor programming skills, 
while some others seem intentional: either should discourage users to join.
However, to transmit a feeling of security, contract owners shelter themselves
behind the motto that \emph{the code is publicly accessible},
assuming that everyone can read it and assess its reliability.
Since bugs are often missed even by their own creators, it is hard
to imagine that the average user can read a contract and fully
understand what it really does and what harms can be hidden behind.
Differently from the notorious vulnerabilities which affected the DAO~\cite{DAO}
and the Parity wallet~\cite{parity17jul,parity17nov},
which caused money losses in the order of hundreds of millions of dollars,
the vulnerabilities discussed here involve smaller contracts:
indeed, the vast majority of the contracts in our collection 
stays in less than 100 lines of Solidity code (for comparison, the DAO was $\sim$1200 lines).

To counteract these vulnerabilities, researchers have started  
to develop tools for automatically analysing Ethereum  
contracts~\cite{Luu16ccs,Grishchenko18cav,Mythril18,Tikhomirov18wetseb,Tsankov18ccs}.
These tools manage to detect several common vulnerabilities, 
even though the Turing-completeness of EVM and Solidity make verification unfeasible,
in general.
A parallel line of research is the development of \emph{domain-specific} languages 
for smart contracts (possibly, not Turing-complete),
which can help to improve the precision of analysis techniques,
by reducing the distance between contract specification and implementation.
Several domain-specific languages for smart contracts have been proposed, 
not only targeted to the Ethereum platform.
Among them, FSolidM~\cite{Mavridou18fc} models contracts as finite automata,
which can be translated into Solidity code;
the works~\cite{Biryukov17wtsc,EgelundMuller17bise,Thompson18isola} 
develop languages to specify financial contracts 
in the style of Peyton Jones et al.~\cite{PeytonJones00icfp}.
% Pyramid~\cite{Burge18pyramid}

Our recommendation for users is to apply these tools to verify, 
at least, that the contract they want to join does not suffer from 
vulnerabilities like those discussed in~\Cref{sec:source}.
While this alone does not guarantee that the scheme is fair,
it can guarantee \eg that the contract owner does not 
surreptitiously steal funds.
Domain-specific languages might allow for more sophisticated analyses,
which ideally could verify that the distribution of funds among users is fair%
\footnote{For instance, BitML is an abstract language for Bitcoin contracts, 
which supports the verification of fairness for gambling games, like multi-player lotteries~\cite{BZ19post}.}.

\paragraph{Recommendation \#3: analyse the transaction logs.}
Our analyses in~\Cref{sec:gains-losses,sec:timing,sec:inequality} 
have shown that, despite the many peculiarities, 
the transaction logs of Ponzi schemes seem to share some general patterns:
\begin{inlinelist}
\item only a few users have a ratio greater than $1$:
the most numerous classes are those of users who never received any money back, 
or have a ratio between $0$ and $1$;

\item most Ponzi schemes have a relatively short lifespan,
consisting in a peak of intense activity followed by a period of stagnation;

\item the Gini coefficients of the \emph{payouts} of Ponzi schemes 
  tend to be high (more than 80\% in our collection),
  meaning a strong inequality in the distribution of money. 

\end{inlinelist}
Even though none of these features alone seems enough to separate Ponzi schemes
from other contracts, these features can be used together to train classifiers
which \emph{automatically} detect Ponzi schemes.
Along these lines, a preliminary version of our dataset has been used in~\cite{Chen18www,Chen19access}
to experiment with learning strategies to classify Ethereum Ponzi schemes. 
The classifier in~\cite{Chen18www,Chen19access} uses simple features of transaction logs
(\eg, number of payments, contract balance, proportion of investors who received at least one payment, \etc), 
as well as features of the contracts EVM code (\eg, the number of occurrences of certain opcodes).
The measurements in~\cite{Chen18www,Chen19access} show that code features are more discriminating than
transaction features --- somehow counter-intuitively, 
since EVM features do not seem to carry any insight on the nature of the contract.
However, this discrepancy may be due to an over-simplification in the choice of transaction features:
using more sophisticated features, inspired to those discussed in~\Cref{sec:timing,sec:gains-losses},
may help improve the precision of the classification.
The analysis techniques of Ponzi schemes in Bitcoin~\cite{BPS18cvcbt,Toyoda17globecom} demonstrate that 
the automatic classification of Ponzi schemes from the transaction history alone is feasible with a high level of accuracy.

\paragraph{Future works.}
The Ponzi schemes we have presented in this paper can be seen as the first wave of Ethereum-based scams.
In a preliminary version of this paper that we put online on March 10th, 2017
\footnote{\href{https://arxiv.org/abs/1703.03779v1}{\code{arxiv.org/abs/1703.03779v1}}.},
we had foreseen a second wave of scams:
% as Ethereum consolidates its position as a platform for smart contracts and as a cryptocurrency, 
% criminals will exploit it to host their scams.
\begin{center}
\begin{minipage}{0.97\textwidth}
\say{{\it%
very likely they will be less recognizable as such than the ones collected in this survey. 
For instance, they could mix multi-level marketing, token sales, and games,
to realize complex smart contracts, 
which would be very hard to correctly classify as Ponzi schemes or legit investments}}
\end{minipage}
\end{center}

We believe that this expectation may have come true
with Initial Coin Offerings (ICOs), a means of crowdfunding based on the trade of crypto-tokens,
through which more than $3 \USD$ billions have been collected in 2017%
\footnote{Source: \href{https://www.coinschedule.com/stats.html}{\code{www.coinschedule.com/stats.html}}.},
as well as crypto-collectibles games like \emph{CryptoKitties} and its followers.
The absence of specific regulations in Europe and in the US,
and the general difficulty of governing decentralised cryptocurrencies,
have made these schemes attractive also for scammers:
indeed, a few ICOs have been unmasked as Ponzi schemes by financial authorities~\cite{Shin17forbes,Zetzsche17ico}.
A relevant research line for future works could be that of studying these kinds of ``pseudo'' Ponzi schemes,
which share many similarities with Ponzis, 
although failing to meet the requirements we have specified in~\Cref{sec:methodology}.
Some features which apply well to ``pure'' Ponzi schemes, like \eg the gain ratio,
seem appropriate also to characterize these ``pseudo'' Ponzi.

\section*{References}
\bibliography{main}

\iftoggle{draft}{}{
\appendix
\section{Appendix: general charts}
\label{app:charts}

In this section, we gather the charts for a selection of $23$ contracts,
chosen among those with the most interesting features: high volume of
payments or  high number of users, or high number of transactions.

\definecolor{col1}{RGB}{20, 20, 255}
\definecolor{col2}{RGB}{136, 136, 136}
\pgfplotsset{compat=newest}
\pgfplotsset{major grid style={gray!50}}

\setcounter{gainListCounter}{0}

\foreach \y in \gainList 
{
\stepcounter{gainListCounter}
%printing all the files in gainList
\begin{landscape}
\begin{figure}[!htbp]
\centering
\scriptsize
\foreach \x in \y  
{
\begin{tikzpicture}
\begin{axis}[
    width  = 0.5\linewidth, 
    %nodes near coords, %writes the y numbers 
    xtick=data, xticklabels = \empty,
    height = 5cm,  
    xmajorgrids = true, ymajorgrids = true,
    xlabel absolute, xlabel style={yshift=-0.5cm},
    y tick label style={font=\tiny,major tick length=0pt},
    enlargelimits = false,   scaled y ticks = false,
    extra y tick style={grid=major, tick label style={xshift=-1cm}},
    legend style={ ultra thin}, legend pos=north west,  
]
\legend{\x}
\pgfplotstableread[col sep=comma]{results/gainLoss__\x.csv}\data
\addplot [color=col1, thick, mark=none] table[x index = {0}, y index = {1}]{\data};
\end{axis}
\end{tikzpicture} 
}
\caption{Gain in ether (on y-axis) per user (on x-axis), per
  contract. Users have been ordered by increasing values. }
\label{fig:gain-charts:\thegainListCounter}
\end{figure}
\end{landscape}
}

\setcounter{gainListCounter}{0}

\foreach \y in \gainList 
{
\stepcounter{gainListCounter}
%printing all the files in ratioList
\begin{figure}[!htbp]
\centering
\scriptsize
\foreach \x in \y  
{
\begin{tikzpicture}
  \begin{axis}[ 
    height = 5.5cm, 
    ybar, ymin=0,
    flexible xticklabels from table={results/gainRatio__\x.csv}{label}{col sep=comma},
    xticklabel style={font=\tiny}, y tick label style={font=\tiny}, 
    xtick=data,
    legend style={ultra thin, fill=none}, legend pos=north east,  
    nodes near coords, nodes near coords align={vertical}
    ]
    \pgfplotstableread[col sep=comma]{results/gainRatio__\x.csv}\datatable
    \legend{\x}
    \addplot table[x expr=\coordindex,y=value]{\datatable};
  \end{axis}
\end{tikzpicture}
}
\caption{Number of users grouped by gain-ratio. 
  Label \emph{0} means no money has been received; 
  label \emph{1} indicates a ratio between $0$(strict) and
  $1$(included); 
  label \emph{2} indicates a ratio between $1$(strict)
  and $2$(included), and so on.  
  Label \emph{$\infty$} indicates users who  
  have never sent anything but have received something; 
  and label \emph{other} indicates the rest of the users.}
\label{fig:ratio-charts:\thegainListCounter}
\end{figure}
}

\definecolor{blue}{RGB}{20, 20, 255}
\definecolor{red}{RGB}{255, 20, 20}

\setcounter{gainListCounter}{0}

\foreach \y in \gainList 
{
\stepcounter{gainListCounter}
%printing all the files in gainList
\begin{figure}[!htbp]
\centering
\scriptsize
\foreach \x in \y  
{
\begin{tikzpicture}
\begin{axis}[ 
       width  = 1\linewidth, height = 3cm,
       date coordinates in=x, table/col sep=comma,
       xticklabel=\tiny \day-\month, 
       legend style={ ultra thin}, legend pos=north east,  
       xticklabel style={rotate=90, anchor=near xticklabel}]
   \legend{\x}
   \pgfplotstableread[col sep=comma]{results/dateInFlow__\x.csv}\data
   \addplot+[only marks,mark size=0.5, draw=blue] table[x index = {0}, y index = {1}] \data; 
   \pgfplotstableread[col sep=comma]{results/dateOutFlow__\x.csv}\data
   \addplot+[only marks,mark size=0.5, draw=red] table[x index = {0}, y index = {1}] \data;
\end{axis}
\end{tikzpicture}  
}
\caption{Inflow (blue) and outflow (red) timing: on x axis the time of
  transaction (day-month); on the y axis the amount of ether.}
\label{fig:timing-charts:\thegainListCounter}
\end{figure}
}

\begin{figure}
\small
\centering
\setlength{\extrarowheight}{2pt}
\renewcommand{\arraystretch}{1.2}
\tabcolsep=0.11cm
\begin{tabular}{ |c c c c || c | } 
\hline
 \thead{\bf Contract \\ }&\thead{\bf Invested \\  } & \thead{\bf Received \\ } & \thead{\bf Net gain \\ } &\thead{\bf Ratio \\ }\\	
\hline \hline
Government	&	109	&	1099	&	990	&	10	\\
\hline
\multirow{2}{*}{EthereumPyramid}&	0	&	90.8	&	90.8	&	$\infty$	\\
	&  	0	&	226.8	&	226.8	&	$\infty$	\\
\hline
\multirow{3}{*}{ProtectTheCastle}&	0.2	&	64	&	63.8	&	320	\\
	&	0.01	&	0.7	&	0.6	&	70	\\
	&	0.8	&	16.3	&	15.5	&	19.4	\\
\hline
TreasureChest	&	0.1	&	6.2	&	6.1	&	62	\\
\hline
ZeroPonzi	&	495	&	618.7	&	123.7	&	1.2	\\
\hline
\multirow{5}{*}{Doubler2}&	1	&	486.8	&	485.8	&	486.8	\\
	&	370.6	&	654.5	&	283.9	&	1.7	\\
	&	200	&	360	&	160	&	1.8	\\
	&	254	&	457.2	&	203.2	&	1.8	\\
	&	601	&	1081.8	&	480.8	&	1.7	\\
\hline
\multirow{5}{*}{DynamicPyramid}	&	0.9	&	411.6	&	410.7	&	457.3	\\
	&	362	&	543.5	&	181.5	&	1.5	\\
	&	244	&	366	&	122	&	1.5	\\
	&	212	&	318	&	106	&	1.5	\\
	&	810	&	1015.5	&	205.5	&	1.2	\\
\hline
Etheramid1	&	0	&	30.6	&	30.6	&	$\infty$\\   
\hline
\multirow{3}{*}{Doubler}&	1	&	55.4	&	54.4	&	55.4	\\
	&	450	&	784	&	334	&	1.7	\\
	&	720.2	&	1013.9	&	293.6	&	1.4	\\
\hline
Thesimplegame	&	0	&	17.7	&	17.7	&	$\infty$	\\
\hline
Doubler3	&	0	&	40.2	&	40.2	&	$\infty$	\\
\hline
Quick1	&	0	&	10.9	&	10.9	&	$\infty$	\\
\hline
\multirow{2}{*}{Myscheme}&	1	&	17.1	&	16.1	&	17.17	\\
 	&	2	&	55.1	&	53.1	&	27.5	\\
\hline
Bunny	&	0	&	1.2	&	1.2	&	$\infty$	\\
\hline
DoubleTx	&	1.2	&	39.5	&	38.3	&	32.9	\\
\hline
Multi33v	&	0	&	1.5	&	1.59	&	$\infty$	\\
\hline
DianaEthereum-x18	&	4.6	&	539.1	&	534.4	&	115.3	\\
\hline
\multirow{6}{*}{Rubixi}	&	0	&	4.5	&	4.5	&	$\infty$	\\
	&	0	&	0.03	&	0.03	&	$\infty$	\\
	&	0	&	8.3	&	8.3	&	$\infty$	\\
	&	0	&	8.00E-18	&	8.00E-18	&	$\infty$	\\
	&	0	&	1.1	&	1.1	&	$\infty$	\\
	&	0	&	0.5	&	0.5	&	$\infty$	\\ 
\hline
\end{tabular} 
\caption{Details (in ETH) for those users with gain ratio $>10$ or
  equals to \emph{$\infty$} or with net gain $>100$, per contracts. On
  column 1, there is the name of the contract the user joined; then
  how much he invested; how much he received; then the difference
  between column 3 and column 2, and then the ratio between column 3 and
  column 2. Values have been cut to the first non-zero decimal digit.}
\label{fig:highest-gains}
\end{figure}

\begin{figure}
\small
\centering
\setlength{\extrarowheight}{5pt}
\renewcommand{\arraystretch}{1.2}
\tabcolsep=0.11cm
\begin{tabular}{ |c c c c| } 
\hline
 \thead{\bf Contract} &   \thead{ \bf Users joining \\ the scheme} &  \thead{\bf Users \emph{not} \\  gaining}   &    \thead{\bf Percent} \\ 
\hline
Doubler&	92&	81&	88\%	\\
ShinySquirrels1&	23&	20&	87\%	\\
GreedPit&	61&	52&	85\%	\\
Doubler3&	18&	15&	83\%	\\
Rubixi&	99&	82&	83\%	\\
Etheramid1&	103&	84&	82\%	\\
\hline
Doubler2&	210&	165&	79\%	\\
Quick1&	13&	10&	77\%	\\
DynamicPyramid&	174&	133&	76\%	\\
Government&	40&	30&	75\%	\\
TreasureChest&	16&	12&	75\%	\\
Newponzi&	19&	14&	74\%	\\
ProtectTheCastle&	98&	72&	73\%	\\
Ethstick&	45&	33&	73\%	\\
DoubleTx&	22&	16&	73\%	\\  
Thesimplegame&	11&	8&	73\%	\\
\hline
EthereumPyramid&	326&	221&	68\%	\\
ZeroPonzi&	46&	27&	59\%	\\
Bunny&	58&	33&	57\%	\\
DianaEthereum-x18&	129&	66&	51\%	\\
Multi33v&	14&	7&	50\%	\\
LittleCactus&	50&	24&	48\%	\\
Myscheme&	52&	23&	44\%	\\
\hline
\end{tabular} 
\caption{Percent of users which have not gained anything from joining
  the scheme, per contract. Column 2 counts the users who entered the
  scheme by sending some money, and column 3 counts how many of them
  have gain ratio less or equal to 1. }
\label{fig:not-gaining}
\end{figure}

}

\end{document}